\documentclass{pasa}%
\usepackage{lscape, morefloats, graphicx}
\title{Analysis of the disc components of our Galaxy via kinematic and spectroscopic procedures}
\author[Karaali et al.]{S. Karaali$^1$\thanks{karsa@istanbul.edu.tr}, S. Bilir$^1$, E. Yaz G\"ok\c ce$^{1,2}$, and O. Plevne$^3$\\
     \affil{$^1$Istanbul University, Faculty of Science, Department of Astronomy and Space Sciences, 34119, Beyaz\i t, Istanbul, Turkey}
     \affil{$^2$Ege University, Faculty of Science, Department of Astronomy and Space Sciences, 35100 Bornova, Izmir, Turkey}
     \affil{$^3$Istanbul University, Institute of Graduate Studies in Science, Programme of Astronomy and Space Sciences, 34116, Beyaz{\i}t, Istanbul, Turkey\\}
}
\jid{PASA}
\doi{10.1017/pas.\the\year.xxx}
\jyear{\the\year}
\usepackage{natbib}
\usepackage{graphicx}
\graphicspath{%
    {converted_graphics/}% inserted by PCTeX
    {/}% inserted by PCTeX
}
\begin{document}
\begin{abstract}
We used the spectroscopic and astrometric data provided from the GALAH DR2 and {\it Gaia} DR2, respectively, for a large sample of stars to investigate the behaviour of the [$\alpha$/Fe] abundances via two procedures, i.e. kinematically and spectroscopically. With the kinematical procedure, we investigated the distribution of the [$\alpha$/Fe] abundances into the high/low probability thin disc, and high/low probability thick-disc populations in terms of total space velocity, [Fe/H] abundance, and age. The high probability thin-disc stars dominate in all sub-intervals of [$\alpha$/Fe], including the rich ones: [$\alpha$/Fe]$>0.3$ dex, where the high probability thick-disc stars are expected to dominate. This result can be explained by the limiting apparent magnitude of the GALAH DR2 ($V<14$ mag) and intermediate Galactic latitude of the star sample. Stars in the four populations share equivalent [$\alpha$/Fe] and [Fe/H] abundances, total space velocities and ages. Hence, none of these parameters can be used alone for separation of a sample of stars into different populations. High probability thin-disc stars with abundance $-1.3<{\rm[Fe/H]}\leq -0.5$ dex and age $9<\tau\leq13$ Gyr are assumed to have different birth places relative to the metal rich and younger ones. With the spectroscopic procedure, we separated the sample stars into $\alpha$-rich and $\alpha$-poor categories by means of their ages as well as their [$\alpha$/Fe] and [Fe/H] abundances.  Stars older than 8 Gyr are richer in [$\alpha$/Fe] than the younger ones. We could estimate the abundance [$\alpha$/Fe]=0.14 dex as the boundery separating the $\alpha$-rich and $\alpha$-poor sub-samples in the [$\alpha$/Fe]$\times$[Fe/H] plane.   
\end{abstract}
\maketitle%

\section{Introduction}
Different procedures and parameters are used to understand the formation and evolution of our Galaxy. The iron element relative to the hydrogen, [Fe/H], was used for a long time as the main parameter, and still it is used in the new procedures for this purpose. Discovery of the metal-poor stars leds the separation of the stars in our Galaxy into disc (metal-rich) and halo (metal-poor) populations which occupy the short and large vertical distances of our Galaxy, respectively. The need of a third population for fitting the space densities observed by \cite{Gilmore83} increased the populations of our Galaxy from two to three. The new population covers the stars with intermediate metallicities and vertical distances. Thus, the disc component of our Galaxy is separated into two sub-components, thin disc and thick disc. The investigation of the formation and evolution of our Galaxy is usually based on these components, thin disc, thick disc and halo. The ages of the three mentioned Galactic components are young, intermediate, and old, in the order just given.

The main source of the iron-peak elements is the Type-Ia supernovae which are produced in a few Gyr. There is another set of elements, alpha ($\alpha$) elements, which are mostly produced from the Type II supernovae in about 20 Myr \citep{Wyse88}. Due to difference between the production times of the iron peak and $\alpha$-elements, one expects that old stars are $\alpha$-rich but iron-poor, while the young ones are iron-rich and hence $\alpha$-poor relative to the iron abundance, i.e. [$\alpha$/Fe]. 

The formation and evolution of our Galaxy can be investigated kinematically as well. The procedure used is the combination of the probability function $f(U,V,W)$ and the observed fractions ($X$) of each population in the solar neighbourhood as explained in \citet{Bensby03, Bensby05}:
%Eq.1
\begin{eqnarray}
\scriptstyle f(U,V,W) = k\times\exp[-(U_{LSR}^2/2\sigma_{U}^2)-((V_{LSR}-V{asym})^2/2\sigma_{V}^2)\\ \nonumber
\scriptstyle-(W_{LSR}^2/2\sigma_{W}^2)]
\end{eqnarray}
where it is assumed that the Galactic space velocity components, $U_{LSR}$, $V_{LSR}$, and $W_{LSR}$ of the stellar populations in the thin disc, thick disc, and halo have Gaussian distributions, $k=[(2\pi)^{3/2}\sigma_U \sigma_V \sigma_W]^{-1}$ normalizes the expression, $\sigma_U$, $\sigma_V$, and $\sigma_W$ are the characteristic velocity dispersions, and $V_{asym}$ is the asymmetric drift. Thus, one gets two relative probabilities for a star as in the following:

% Eq.2 
\begin{eqnarray}
TD/D=(X_{TD}/X_D)\times(f_{TD}/f_D),\\ \nonumber
TD/H = (X_{TD}/X_H)\times(f_{TD}/f_H)  
\end{eqnarray}
where $X_D$, $X_{TD}$, and $X_H$ refer to the observed local fractions for the thin disc, thick disc, and halo, respectively. If $TD/D$ is equal to a big number, the probability of the star in question of being thick disc star relative to the thin disc is relatively high. For example, in the case of $TD/D=10$, the probability of a star belonging to the thick disc component is 10 times bigger than the thin disc one. In the opposite case, i.e. if $TD/D$ is a small number, say $1/10$, the star is probably a thin disc star. A similar case holds for the relative probability $TD/H$.

The $\alpha$-elements have been phenomena of Galactic research since the beginning of this century. Many studies are based on the distribution of the $\alpha$-elements relative to the iron abundance, [$\alpha$/Fe]$\times$[Fe/H]. Also, the population types of the stars, estimated kinematically, are usually indicated by different symbols on this diagram. One can see the following picture on this diagram: The halo population covers the [Fe/H]-poor and $\alpha$-rich stars, and the thick disc population is dominant with intermediate [Fe/H] and $\alpha$ stars, while the thin- disc stars are [Fe/H]-rich and $\alpha$-poor. These studies can be found in \cite{Stephens02}, \cite{Allende04}, \cite{Bensby03}, \cite{Bensby14}, \cite{Reddy06}, \cite{Nissen10}, \cite{Adibekyan12}, \cite{Bovy12}, \cite[][and references therein]{Haywood13}. The largest set of $\alpha$-elements is the Hypatia catalogue \citep{Hinkel14}. Alpha abundances can be found also in large surveys, such as RAVE \citep{Steinmetz06}, APOGEE \citep{Allende08}, SEGUE \citep{Yanny09}, LAMOST \citep{Zhao12}, GES \citep{Gilmore12}, GALAH \citep{Heijmans12}, and GIBS \citep{Zoccali14}.               

Although there is a large improvement in attribution of spectroscopic and kinematic data to a given star population, there is still some problems to be solved. For example, \citet{Haywood13} and \citet{Hayden17} showed that the metal-rich intermediate stars have kinematics similar to the thin disc. \citet{Bland-Hawthorn19} used a different terminology for the two discs of our Galaxy, i.e. $\alpha$-rich disc and $\alpha$-poor disc, instead of thick disc and thin disc, respectively. The flaring $\alpha$-poor disc in their Fig. 1 indicates a longer scale-length for the $\alpha$-poor disc relative to the $\alpha$-rich disc. The simple choice of the boundary between $\alpha$-rich disc stars and $\alpha$-poor disc stars in the [$\alpha$/Fe]$\times$[Fe/H] diagram, in \citet{Bland-Hawthorn19}, is carried out dynamically where they showed that the actions of the two star categories distinct. Another recent work is the study of \citet{Buder19}. These authors first showed that stars of the high-$\alpha$ sequence are older ($>$ 8 Gyr) than stars in the low-$\alpha$ sequence with the iron abundances $-0.7<{\rm [Fe/H]}<+0.5$ dex, and then they applied their finding to the metal-rich stars which become indistinguishable in the [$\alpha$/Fe]$\times$[Fe/H] diagram. Also, \citet{Buder19} found a continuous evolution in the high-$\alpha$ sequence up to super-solar [Fe/H].

The most recent study that we would like to cite is the one of \citet{Haywood19}. These authors argue that the Galactic disc within 10 kpc has been enriched to solar metallicity by the thick disc. Then, two different enrichments took place: The inner disc, $R<6$ kpc, continued enrichment after a quenching phase (7-10 Gyr), while at larger distances radial flows of gas diluted the metals left by the thick disc formation. \citet{Haywood19} stated that the strong correlation between ages, $\alpha$-abundances, and metallicities for the inner disc stars is due to the closed-box type evolution of the inner disc and bulge. One can see a good fit of the mentioned data to their model. However, the same case does not hold for the outer disc stars, i.e. mono-abundance populations are not mono-age populations, because stars with a given age cover a large range in metallicity. As stated by the authors, the birth radius of the stars should be considered to obtain a mono-age population. 

The spectroscopic data have the advantage of representing the structure of the gas cloud they were formed, while kinematic data  may change by time due to the effect of some structures of our Galaxy, such as the central bar, spiral arms, aggregation material, molecular clouds etc. From the other hand, it has been shown in the recent studies that there is an overlap between the abundances of metals observed for stars in different populations i.e. the iron metallicity of the old thin-disc stars is compatible with the one of the thick-disc stars \citep[cf.][]{Haywood13}. Then, one can deduce that the spectroscopic and kinematic parameters should be combined with age for a better understanding of the formation and evolution of the Galactic components, i.e. thin disc, thick disc, and halo. 

In this study, we used the spectroscopic and astrometric data provided from the GALactic Archaeology with HERMES (GALAH) Data Release (DR2) \citep{Buder18} and {\it Gaia} DR2 from the European Space Agency mission {\it Gaia} \citep{Gaia16, Gaia18}, respectively, for a large sample of F and G type stars for two purposes: 1) to investigate the distribution of their [$\alpha$/Fe] abundances into the thin and thick-disc populations in terms of total space velocity, [Fe/H] abundance and age, and 2) to separate them into two categories such as $\alpha$-rich and $\alpha$-poor stars with respect to [Fe/H] abundance by using the age parameter, but avoiding any kinematical/dynamical parameter. GALAH is a large-scale stellar spectroscopic survey of our Galaxy and designed to deliver chemical information complementary to a large number of stars covered by the {\it Gaia} mission. GALAH DR2 \citep{Buder18} contains 342682 stars provided with stellar parameters and abundances for up to 23 elements. The High Efficiency and Resolution Multi-Element Spectrograph (HERMES) mounted on the Anglo-Australian Telescope provides multi-object ($n \approx 392$) and high resolution ($R\approx 28000$). 

{\it Gaia} relies on the proven principles of ESA's {\it Hipparcos} mission to help solve one of the most difficult yet deeply fundamental challenges in modern astronomy: the creation of an extraordinarily precise three-dimensional map of about one billion stars throughout our Galaxy and beyond. The {\it Gaia} mission delivers an astronomical catalogue and data archive of unprecedented scope, accuracy and completeness. Combined with the additional of precise astrometric information from the {\it Gaia} satellite one has the ability to determine the precise kinematic, dynamic, and temporal structure of stellar populations throughout the Galaxy. We organized the paper as follows. \S 2 is devoted to the star sample, investigation of the sample stars via kinematic and spectroscopic procedures is given in \S 3, and finally a summary and discussion is presented in \S 4.

% Figure 1
\begin{figure}[b]
\centering
\includegraphics[scale=0.6, angle=0]{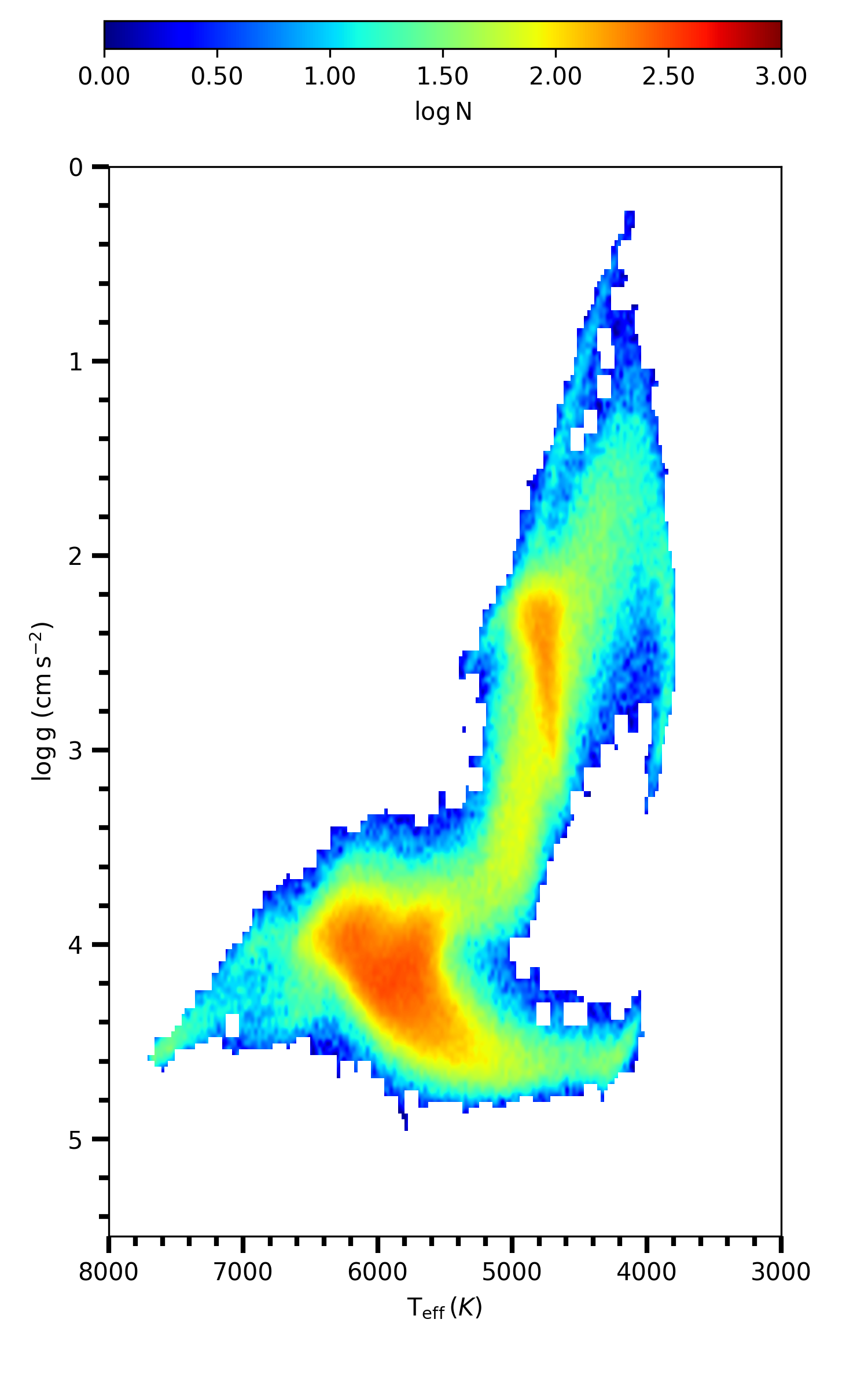}
\caption{$\log g \times T_{eff}$ diagram of the stars in the GALAH DR2 with colour-coded for the number of stars.}
\label{fig1}
\end{figure} 

% Figure 2
\begin{figure*}[t]
\centering
\includegraphics[width=16cm,height=7cm, keepaspectratio]{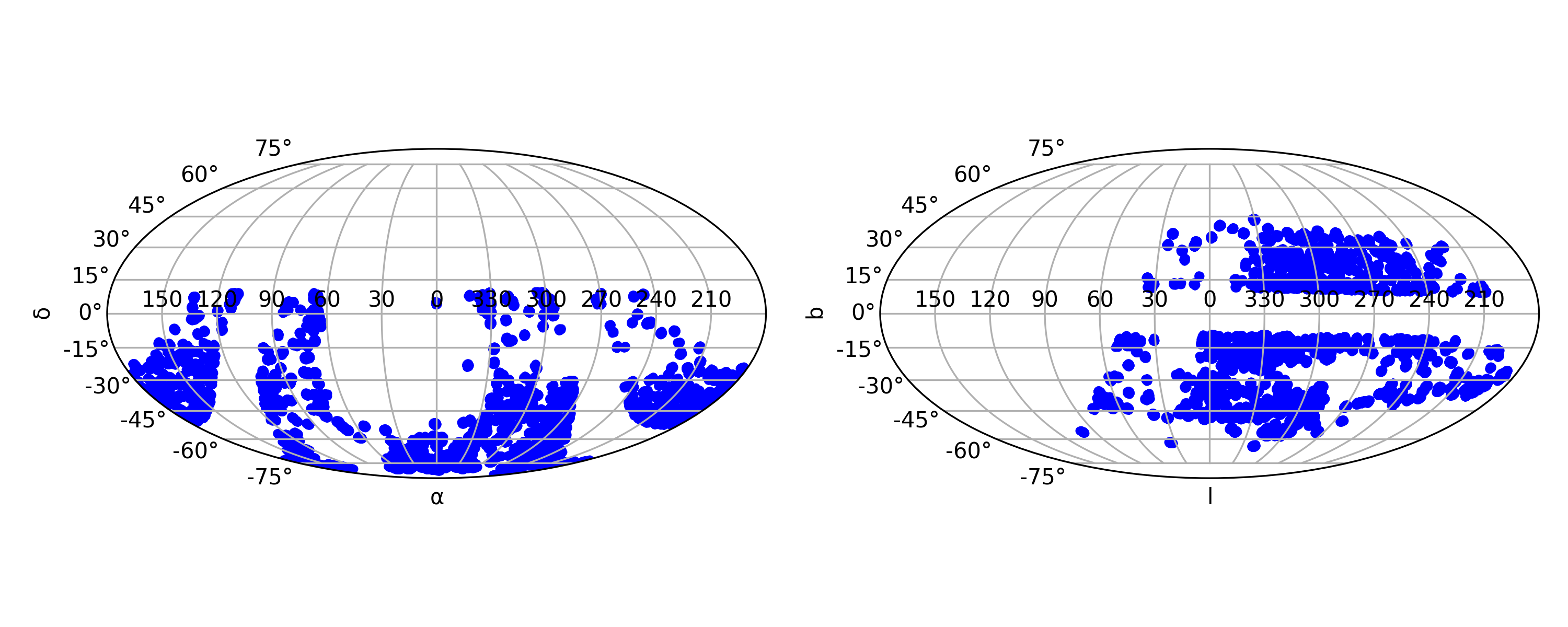}
\caption{Distributions of the stars in the GALAH DR2 in the equatorial coordinate (left) and Galactic coordinate systems (right).}
\label{fig2}
\end{figure*} 

\section{Star Sample}
The star sample is provided from the GALAH DR2 survey \citep{Buder18} which covers the spectral data of 342682 stars. We applied the following constrains to obtain a star sample of F-G type dwarf stars with accurate trigonometric parallaxes: 5250 $\leq T_{eff} \leq$ 7160 K, $\log g \geq 4$ (cm s$^{-2}$), $\sigma_\varpi/\varpi \leq 0.05$, and $G\leq14$ mag. The 5\% trigonometric parallax uncertainties minimize bias. The $\log g \times T_{eff}$ diagram for all stars and the distribution of the 69157 stars in equatorial and Galactic coordinate systems are shown in Fig. \ref{fig1} and Fig. \ref{fig2}, respectively. [$\alpha/$Fe] and [Fe/H] abundances as well as the radial velocities are provided from the GALAH DR2 catalogue \citep{Buder18}, while the proper motions and trigonometric parallaxes, used in distance estimations, are taken from the {\it Gaia} DR2 \citep{Gaia18}. Radial velocities, proper motions and distances are combined to obtain space velocity components ($U$, $V$, $W$) of the sample stars which are calculated with \cite{Johnson87} standard algorithms and transformation matrices of a right-handed system epoch J2000. Hence, $U$, $V$, and $W$ are the space velocity components of a star with respect to the Sun, where $U$ is positive towards the Galactic centre, $V$ is positive in the direction of the Galactic rotation, and $W$ is positive towards the North Galactic Pole. We applied the procedure of \citet{Mihalas81} to the distribution of the sample stars and estimated the first-order of Galactic differential rotation corrections for the $U$ and $V$ velocity components. The $W$ velocity component is not affected by Galactic differential rotation. The ranges of these corrections are $-69<dU<29$ and $-4<dV<5$ km s$^{-1}$ for $U$ and $V$, respectively, while their means are $\langle dU\rangle=-8.53$ and $\langle dV \rangle=0.43$ km s$^{-1}$. The Galactic rotational velocity of the Sun is adopted as 222.5 km s$^{-1}$ \citep{Schonrich10}. Then, the space velocity components are reduced by applying the solar local standard of rest (LSR) values of \citet{Coskun11} for all sample stars ($U$, $V$, $W$)$_{LSR}$=($8.83\pm0.24$, $14.19\pm0.34$, $6.57\pm0.21$) km s$^{-1}$. The numerical values of the characteristic velocity dispersions, $\sigma_U$, $\sigma_V$, and $\sigma_W$ and the asymmetric drift $V_{asym}$ used in Eq. (1) for our sample stars are provided from \citet{Bensby03}, while the observed local fractions for the thin disc, thick disc, and halo, $X_D$, $X_{TD}$, and $X_H$ used in Eq. (2) are taken from \cite{Buser99}, and listed in Table \ref{table:1}.

%Table 1
\begin{table}
\centering
\setlength{\tabcolsep}{3pt}
\caption{Numerical values of the characteristic velocity component dispersions, $\sigma_U$, $\sigma_V$, and $\sigma_W$, the asymmetric drift $V_{asym}$ and the observed local fractions for the thin disc (D), thick disc (TD), and halo (H).}  
\begin{tabular}{lrcccc}
\hline
Component & $X$& $\sigma_U$ &  $\sigma_V$ & $\sigma_W$  & $V_{asym}$\\
& & \multicolumn{4}{c}{(km s$^{-1}$)} \\
\hline
Thin disc (D)  & 0.94	  & 35 & 20 & 16 & -15 \\
Thick disc (TD)& 0.06	  & 67 & 38 & 35 & -46 \\
Halo (H)       & 0.0015   &160 & 90 & 90 & -220\\
\hline
\end{tabular}
\label{table:1}
\end{table}

%Table 2
\begin{table*}
\centering
\setlength{\tabcolsep}{3pt}
\caption{Data for sample stars. The columns give: ID, star, total space velocity, relative population probabilities of the stars, $[\alpha/{\rm Fe}]$ and [Fe/H] abundances, and age.}  
    \begin{tabular}{lccccccccc}
\hline
ID & Star & $S_{LSR}$ & $TD/D$ & $TD/H$& $[\alpha/{\rm Fe}]$ &  [Fe/H] & $\log \tau$ &\\
   &      &(km s$^{-1}$)&         &       &      (dex)                     &  (dex) &   (yr) \\
\hline
    1 & 00000025--7541166 & 32.66$\pm$0.27 & 0.01 & 4735.58 & -0.09$\pm$0.02 &  0.18$\pm$0.06 & 9.85$\pm$0.19 \\
    2 & 00000040--5114382 & 45.61$\pm$0.21 & 0.01 & 4545.41 & -0.04$\pm$0.02 &  0.08$\pm$0.06 & 9.68$\pm$0.20 \\
    3 & 00000133--6500170 & 38.15$\pm$1.24 & 0.01 & 4838.60 & -0.02$\pm$0.03 & -0.10$\pm$0.08 & 9.44$\pm$0.35 \\
    4 & 00000263--7221408 & 47.80$\pm$0.30 & 0.02 & 3763.36 &  0.07$\pm$0.03 & -0.15$\pm$0.09 & 9.86$\pm$0.17 \\
    5 & 00000390--7557323 & 41.94$\pm$0.24 & 0.01 & 4429.91 & -0.06$\pm$0.03 &  0.44$\pm$0.07 & 9.76$\pm$0.20 \\
.. & ..  & ..  & ..  & ..  & ..  & ..  & ..  \\
.. & ..  & ..  & ..  & ..  & ..  & ..  & ..  \\
.. & ..  & ..  & ..  & ..  & ..  & ..  & ..  \\
69153 & 23595092--5110177 & 27.27$\pm$0.18 & 0.01 & 4383.45 &-0.02$\pm$0.03 &  0.10$\pm$0.08 & 9.77$\pm$0.33 \\
69154 & 23595454--7906171 & 74.15$\pm$0.31 & 0.03 & 3286.76 & 0.10$\pm$0.04 & -0.10$\pm$0.09 & 9.87$\pm$0.23 \\
69155 & 23595691--7649056 & 26.57$\pm$0.35 & 0.01 & 5260.63 & 0.10$\pm$0.02 & -0.39$\pm$0.07 & 9.81$\pm$0.34 \\
69156 & 23595693+0504082  & 36.94$\pm$0.38 & 0.01 & 3944.00 & 0.11$\pm$0.03 & -0.27$\pm$0.08 & 9.93$\pm$0.09 \\
69157 & 23595785--7322507 &~~7.05$\pm$0.27 & 0.01 & 5403.93 & 0.07$\pm$0.02 & -0.25$\pm$0.07 & 9.64$\pm$0.28 \\
\hline
    \end{tabular}
  \label{table:2}
\end{table*}

\section{Investigation of the sample stars via kinematic and spectroscopic procedures}
\subsection{Relative probabilities, $[\alpha/{\rm Fe}]$ and [Fe/H] abundances, total space velocities and ages of the sample stars}

% Figure 3
\begin{figure}
\centering
\includegraphics[trim=1cm 1cm 1cm 1cm, scale=0.3, angle=0]{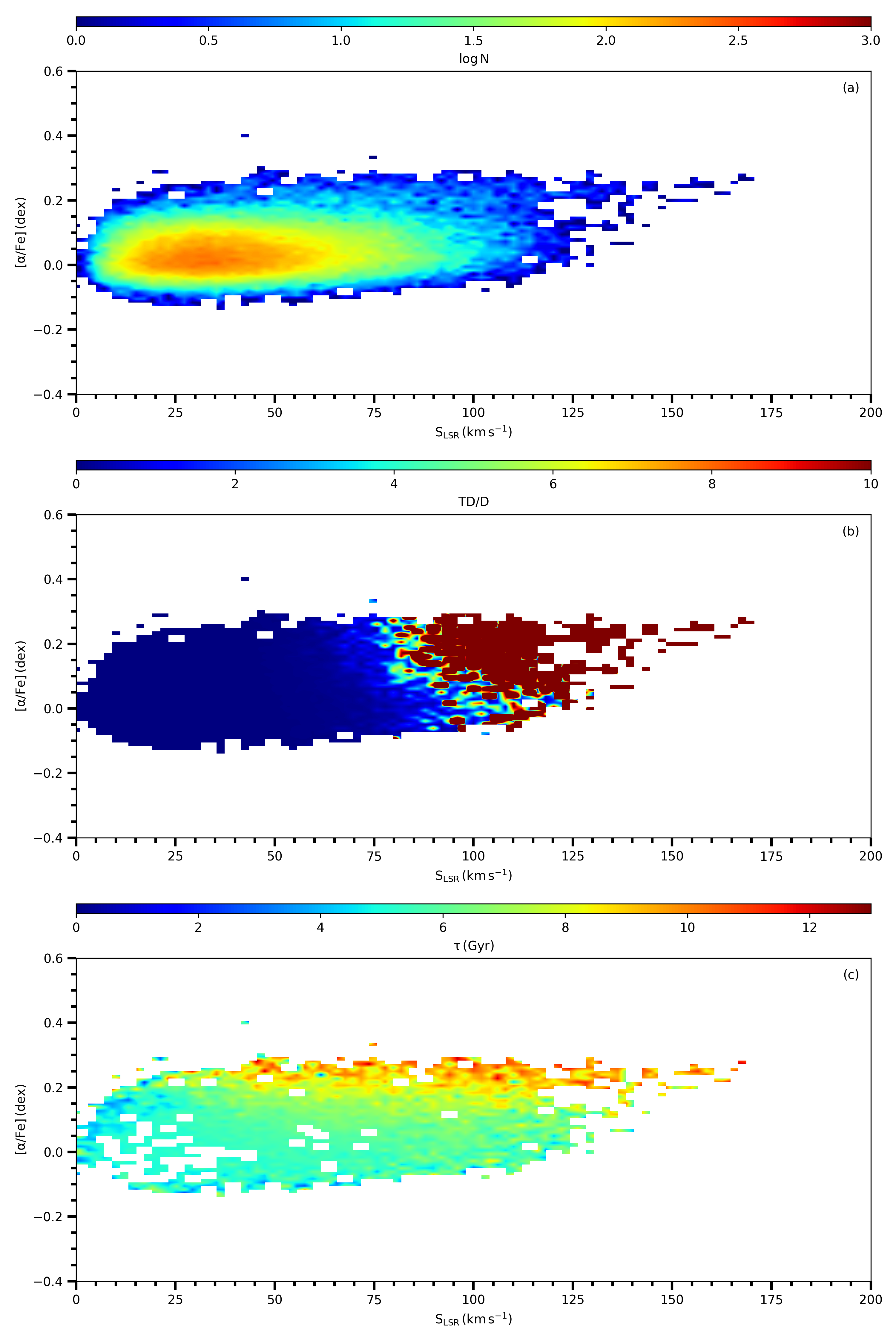}
\caption{Distribution of the $[\alpha/{\rm Fe}]$ abundances against the total space velocities ($S_{LSR}$) in three panels: (a) with colour-coded for the number of stars, (b) with colour-coded for the relative probability ($TD/D$), and (c) with colour-coded for the age ($\tau$).} 
\label{fig3}
\end{figure} 

We used the relative probability functions defined in \S 2, $TD/D$ and $TD/H$, for separation of the sample stars into different populations and determined the $[\alpha/{\rm Fe}]$ abundances of the sample stars in each population as well as their $S_{LSR}$ total space velocities, [Fe/H] abundances and $\tau$ ages (Table \ref{table:2}). The reason of using this procedure is that thin disc stars are $[\alpha/{\rm Fe}]$-poor, [Fe/H]-rich, young and they have relatively small space velocities, while the halo stars are $[\alpha/{\rm Fe}]$-rich, [Fe/H]-poor, old and they have high space velocities; and the numerical values of these parameters for the thick-disc stars are between the numerical values of the two sets of data just mentioned. As noted in \S 2, $[\alpha/{\rm Fe}]$ and [Fe/H] abundances are provided from GALAH DR2, while the ages of the sample stars are calculated by a procedure given in \citet{Onal18}. This procedure is based on Bayesian statistics which includes probability density functions that are obtained from comparative calculations of the theoretical model parameters and the observational parameters \citep{Pont04,Jorgensen05,Duran13}. Following the literature, we separated the sample stars into four populations, i.e. high-probability thin disc: $TD/D\leq 0.1$, low-probability thin disc: $0.1<TD/D\leq 1$, low-probability thick disc: $1<TD/D\leq 10$, and high-probability thick disc: $TD/D>10$. The GALAH probes mainly the thin and thick discs of the Galaxy. Hence, we assume that our sample (statistically) does not cover any halo star. Stars with $0.1<TD/D\leq 10$ are also called ``in-between stars'' \citep{Bensby05} or ``transit stars between D and TD'' in the literature \citep{Adibekyan12}. 

We plotted the [$\alpha/$Fe] abundances against $S_{LSR}$ in three panels of Fig. \ref{fig3}, with colour-coded for the number of stars (a), colour-coded for the relative probability $TD/D$ (b), and colour-coded for the age $\tau$ (c), for a first approximation of the distribution of the sample stars into different sub-samples defined by the parameters just mentioned. Panel (a) shows that the sample stars are dominant in the total space velocity interval of $S_{LSR}<100$ km s$^{-1}$, while one can see a limited number of stars with total velocity as high as $S_{LSR}\approx 175$ km s$^{-1}$. From the other hand, one can deduce from the same panel that most of the sample stars lie in the interval $-0.10<[\alpha/{\rm Fe}]\leq 0.15$ dex, and as the stars with $[\alpha/{\rm Fe}]>0.15$ dex are relatively small in number, we can say that our sample stars are $\alpha$-poor ones. Another result that can be deduced from this panel is that the $[\alpha/{\rm Fe}]$ abundances of the stars with the same total space velocity may be different, one can be $\alpha$-poor, while another one may be relatively a rich star, for example.

The panel (b) of the same figure shows that the $TD/D$ relative probabilities of the stars with relatively high total space velocities, $S_{LSR}>80$ km s$^{-1}$, are relatively high, while the $TD/D$ relative probabilities of the stars with lower total space velocities, $S_{LSR}<80$ km s$^{-1}$, are relatively low. Stars in the second category are also dominant in this panel which indicates that the thin-disc stars are dominant relative to the thick-disc stars, in our sample.

In panel (c), the $\alpha$-abundances of the stars older than 8 Gyr are higher than the ones of younger stars, i.e. $[\alpha/{\rm Fe}]> 0.15$ dex for stars with total space velocities $S_{LSR}>50$ km s$^{-1}$ and $[\alpha/{\rm Fe}]>0.2$ dex for stars with $25<S_{LSR}<50$ km s$^{-1}$. We should emphasize that although the $[\alpha/{\rm Fe}]$ abundances of a group of stars with $S<25$ km s$^{-1}$ are relatively high, $[\alpha/{\rm Fe}]>0.15$ dex, their ages are less than 8 Gyr.

% Figure 4
\begin{figure*}
\centering
\includegraphics[scale=0.3, angle=0]{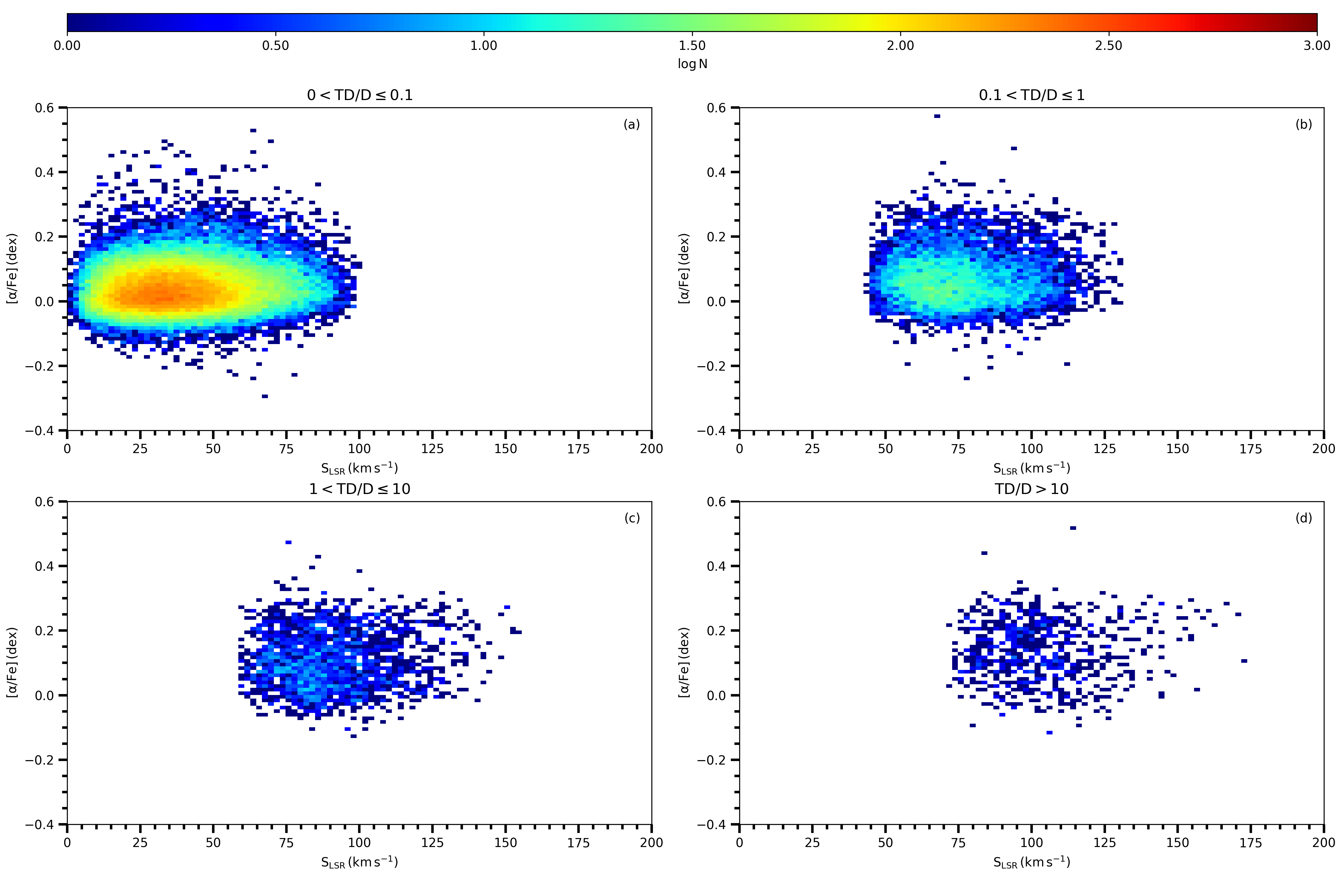}
\caption{Distributions of the $[\alpha/{\rm Fe}]$ abundances against the total space velocities ($S_{LSR}$) for four relative probability functions: $0<TD/D\leq 0.1$, $0.1<TD/D\leq 1$, $1<TD/D\leq 10$, and $TD/D>10$.} 
\label{fig4}
\end{figure*} 

We plotted the $[\alpha/{\rm Fe}]$ abundances versus $S_{LSR}$ velocities for each relative probability function, $0<TD/D\leq 0.1$, $0.1<TD/D \leq 1$, $1<TD/D \leq 10$, and $TD/D>10$ in Fig. \ref{fig4} for a comprehensive investigation of the total space velocity of the sample stars in our study. Fig. \ref{fig4} shows that the number of stars with relative probability $0<TD/D\leq0.1$ in panel (a) are maximum, while they decrease gradually when one goes to the panels (b), (c), and (d), which cover the stars with relative probabilities, $0.1<TD/D \leq 1$, $1<TD/D \leq 10$, and $TD/D>10$, respectively. All stars with $S_{LSR}<45$ km s$^{-1}$ lie in panel (a) and they have relative probabilities $0<TD/D \leq 0.1$. The lower and upper limits of the $S_{LSR}$ velocity in panels (a)-(d) increase gradually. However, there is overlapping between the $S_{LSR}$ velocities of the stars in any two panels, including panel (a) and panel (d). That is, stars can not be separated to the thin disc and thick disc populations by using only their total space velocities. 

The total space velocities, ages and number of the sample stars are also investigated in seven sub-sets of the $[\alpha/{\rm Fe}]$ abundance i.e. $[\alpha/{\rm Fe}]\leq -0.2$, $-0.2<[\alpha/{\rm Fe}]\leq 0$, $0<[\alpha/{\rm Fe}]\leq 0.1$, $0.1<[\alpha/{\rm Fe}]\leq 0.2$, $0.2<[\alpha/{\rm Fe}]\leq 0.3$, $0.3<[\alpha/{\rm Fe}]\leq 0.4$, and $0.4<[\alpha/{\rm Fe}]\leq 0.7$ dex. We omitted discussion of the stars with $[\alpha/{\rm Fe}]\leq -0.2$ dex due to less number of stars, nine stars. Table \ref{table:3} shows that in all sub-intervals of $[\alpha/{\rm Fe}]$, the total space velocity interval of the high-probability thin disc stars ($0<TD/D \leq 0.1$) is limited with $0<S_{LSR}\leq 100$ km s$^{-1}$, while its lower and upper limits shift gradually to higher values when one goes to low-probability thin disc ($0.1<TD/D \leq 1$), low-probability thick disc ($1<TD/D\leq 10$), and high-probability thick disc ($TD/D>10$) stars. Also, the number of high-probability thin disc stars are maximum in all $[\alpha/{\rm Fe}]$ sub-intervals. This is interesting, because one expects the number of high-probability thick disc stars to dominate in the high $[\alpha/{\rm Fe}]$-intervals. However, we should emphasize that the percentage of the high-probability thick disc stars are rather higher in two high $[\alpha/{\rm Fe}]$-intervals, i.e. 28.4$\%$ and 26.6$\%$ in the sub-intervals $0.2<[\alpha/{\rm Fe}]\leq 0.3$ and $0.3<[\alpha/{\rm Fe}]\leq 0.4$ dex, respectively. Another interesting result is that high-probability thin disc stars are dominant in the high abundance ($0.4<[\alpha/{\rm Fe}]\leq 0.7$ dex). Their total space velocities are relatively small, $20.21\leq S_{LSR} \leq 61.32$ km s$^{-1}$, and they are relatively young stars, $3.71 \leq \tau \leq 7.60$ Gyr. However, this unexpected finding can not be generalized for the distribution of the $[\alpha/{\rm Fe}]$ abundances into different populations, but it should be a result of selection effect of the sample stars (see \S 4 for detail).

%TABLE 3
\begin{table*}
\textwidth = 650pt
\setlength{\tabcolsep}{3pt}
\centering
{\scriptsize
\caption{The total space velocities and ages of stars in seven $[\alpha/{\rm Fe}]$ sub-intervals: $[\alpha/{\rm Fe}]\leq -0.2$, $-0.2<[\alpha/{\rm Fe}]\leq 0$, $0<[\alpha/{\rm Fe}]\leq 0.1$, $0.1<[\alpha/{\rm Fe}]\leq 0.2$, $0.2<[\alpha/{\rm Fe}]\leq 0.3$, $0.3<[\alpha/{\rm Fe}]\leq 0.4$, and $0.4<[\alpha/{\rm Fe}]\leq 0.7$ dex in terms of four relative probability intervals defined in the text. The total number of stars in the cited sub-intervals are 9, 17027, 39508, 10312, 2091, 169, and 41, respectively.}
\begin{tabular}{lccc|ccc|ccc|ccc}
\hline
\multicolumn{13}{c}{$[\alpha/{\rm Fe}]\leq-0.2$ dex}\\
\hline
 & \multicolumn{3}{c}{$0<TD/D \leq 0.1$} & \multicolumn{3}{c}{$0.1<TD/D\leq 1$} & \multicolumn{3}{c}{$1<TD/D\leq 10$} & \multicolumn{3}{c}{$TD/D>10$} \\
\hline
$S_{LSR}$-interval & $S_{LSR}$ & $\tau$ & $N$ & $S_{LSR}$ & $\tau$ & $N$ & $S_{LSR}$ & $\tau$ & $N$ & $S_{LSR}$ & $\tau$&$N$ \\
(km s$^{-1}$) & (km s$^{-1}$) & (Gyr) &  & (km s$^{-1}$) & (Gyr) & & (km s$^{-1}$) & (Gyr) &  & (km s$^{-1}$) & (Gyr) &  \\
 \hline
(50,75]  & 57.27 & 6.01 & 7 & $-$ & $-$ & $-$ & $-$ & $-$ & $-$ & $-$ & $-$ & $-$ \\
(75,100] & $-$ & $-$ & $-$ & 81.27 & 5.41 & 2 & $-$ & $-$ & $-$ & $-$ & $-$ & $-$ \\
    Total &  &  & 7    &  &  & 2    &  &  &  &  &  &  \\
     (\%) &  &  & 77.8 &  &  & 22.2 &  &  &  &  &  &  \\
\hline
\multicolumn{13}{c}{$-0.2<[\alpha/{\rm Fe}]\leq 0$ dex}\\
\hline
   (0,25] & 17.39 & 5.10 & 4177 & $-$   & $-$  & $-$ & $-$   & $-$ & $-$ & $-$    & $-$  & $-$  \\
  (25,50] & 36.75 & 5.24 & 8058 & 47.44 & 6.01 & 29  & $-$   & $-$ & $-$ & $-$    & $-$  & $-$  \\
  (50,75] & 58.95 & 5.42 & 2959 & 64.85 & 5.66 & 621 & 68.80 & 5.72 & 21 & $-$    & $-$  & $-$  \\
 (75,100] & 81.94 & 5.51 & 341  & 85.41 & 5.81 & 469 & 86.57 & 5.83 & 145& 89.66  & 5.88 & 22   \\
(100,125] & $-$   & $-$  & $-$  & 106.80& 6.12 & 89  & 107.46& 5.98 & 40 & 112.17 & 5.63 & 35   \\
(125,150] & $-$   & $-$  & $-$  & 128.83& 7.01 & 2   & 131.10& 5.59 & 5  & 130.53 & 5.57 & 13   \\
(150,175] & $-$   & $-$  & $-$  & $-$   & $-$  & $-$ & $-$   & $-$  & $-$& 157.88 & 8.52 & 1    \\
Total &  &  & 15535 &  &  & 1210&  &  & 211  &  &  & 71 \\
(\%)  &  &  & 91.2  &  &  & 7.1 &  &  & 1.2  &  &  & 0.4\\
\hline
\multicolumn{13}{c}{$0<[\alpha/{\rm Fe}]\leq 0.1$ dex}\\
\hline
(0,25]    & 17.62 & 5.11 & 8175 & $-$   & $-$  & $-$ & $-$   & $-$  & $-$ & $-$    & $-$  & $-$ \\
(25,50]   & 36.87 & 5.45 & 17901& 47.86 & 6.07 & 124 & $-$   & $-$  & $-$ & $-$    & $-$  & $-$ \\
(50,75]   & 59.66 & 5.76 & 7392 & 63.78 & 6.13 & 1958& 68.47 & 6.63 & 157 & 73.66  & 7.70 & 5 \\
(75,100]  & 81.99 & 5.99 & 1218 & 85.58 & 6.09 & 1327& 86.09 & 6.40 & 454 & 89.60  & 6.39 & 98 \\
(100,125] & $-$   & $-$  & $-$  & 107.24& 5.99 & 328 & 110.20& 6.18 & 181 & 110.72 & 6.64 & 109 \\
(125,150] & $-$   & $-$  & $-$  & 127.80& 6.41 & 6   & 131.18& 5.99 & 20  & 134.29 & 6.05 & 40 \\
(150,175] & $-$   & $-$  & $-$  & $-$   & $-$  & $-$ & $-$   & $-$  & $-$ & 159.04 & 5.16 & 8 \\
   $>175$ & $-$   & $-$  & $-$  & $-$   & $-$  & $-$ & $-$   & $-$  & $-$ & 222.23 & 5.31 & 7 \\
Total &  &  & 34686 &  &  & 3743&  &  & 812 &  &  & 267 \\
(\%)  &  &  & 87.8  &  &  & 9.5 &  &  & 2.1 &  &  & 0.7 \\
\hline
\multicolumn{13}{c}{$0.1<[\alpha/{\rm Fe}]\leq 0.2$ dex}\\
\hline
   (0,25] & 17.58 & 5.02 & 1620 & $-$   & $-$  & $-$ & $-$   & $-$  & $-$ & $-$   & $-$  & $-$ \\
  (25,50] & 37.32 & 6.00 & 3885 & 47.71 & 7.12 & 48  & $-$   & $-$  & $-$ & $-$   & $-$  & $-$ \\
  (50,75] & 59.72 & 6.56 & 1755 & 63.27 & 7.03 & 892 & 69.17 & 7.46 & 146 & 73.52 & 6.81 & 4 \\
 (75,100] & 81.98 & 6.79 & 324  & 85.02 & 7.15 & 539 & 87.87 & 7.33 & 334 & 90.39 & 7.74 & 146 \\
(100,125] & $-$   & $-$  & $-$  & 107.61& 7.11 & 171 & 109.97& 7.44 & 114 & 110.48& 7.57 & 178 \\
(125,150] & $-$   & $-$  & $-$  & 128.32& 7.21 & 5   & 135.17& 8.31 & 24  & 135.40& 7.60 & 91 \\
(150,175] & $-$   & $-$  & $-$  & $-$   & $-$  & $-$ & 153.88& 8.73 & 2   & 159.67& 8.06 & 24 \\
  $>175$  & $-$   & $-$  & $-$  & $-$   & $-$  & $-$ & $-$   & $-$  & $-$ & 210.67& 7.32 & 10 \\
Total & & & 7584 &  &  & 1655 &  &  & 620 &  &  & 453 \\
(\%)  & & & 73.5 &  &  & 16.0 &  &  & 6.0 &  &  & 4.4 \\
\hline
\multicolumn{13}{c}{$0.2<[\alpha/{\rm Fe}]\leq 0.3$ dex}\\
\hline
(0,25]    & 17.27 & 6.48 & 107 & $-$   & $-$  & $-$ & $-$   & $-$  & $-$ & $-$   & $-$   & $-$ \\
(25,50]   & 38.80 & 7.84 & 333 & 47.73 & 9.92 & 7   & $-$   & $-$  & $-$ & $-$   & $-$   & $-$ \\
(50,75]   & 59.55 & 8.51 & 247 & 64.69 & 8.82 & 215 & 68.86 & 8.73 & 54  & 72.13 & 10.73 & 1 \\
(75,100]  & 82.01 & 8.55 & 44  & 85.21 & 8.68 & 163 & 87.28 & 8.85 & 159 & 90.51 & 9.47  & 113 \\
(100,125] & $-$   & $-$  & $-$ & 108.91& 8.93 & 50  & 110.34& 8.95 & 86  & 111.90& 9.30  & 173 \\
(125,150] & $-$   & $-$  & $-$ & 126.75& 9.29 & 2   & 132.60& 9.16 & 28  & 136.14& 9.40  & 146 \\
(150,175] & $-$   & $-$  & $-$ & $-$   & $-$  & $-$ & 151.13& 6.86 & 2   & 160.74& 9.70  & 94 \\
(75,200]  & $-$   & $-$  & $-$ & $-$   & $-$  & $-$ & $-$   & $-$  & $-$ & 186.47& 9.66  & 36 \\
  $> 200$ & $-$   & $-$  & $-$ & $-$   & $-$  & $-$ & $-$   & $-$  & $-$ & 249.72& 9.72  & 31 \\
Total &  &  & 731  &  &  & 437  &  &  & 329  &  &  & 594 \\
(\%)  &  &  & 35.0 &  &  & 20.9 &  &  & 15.7 &  &  & 28.4\\
\hline
\multicolumn{13}{c}{$0.3<[\alpha/{\rm Fe}]\leq 0.4$ dex}\\
\hline
   (0,25] & 16.84 & 4.87 & 23  & $-$   & $-$  & $-$ & $-$   & $-$  & $-$ & $-$   & $-$   & $-$ \\
  (25,50] & 37.40 & 7.13 & 32  & 47.27 & 6.11 & 1   & $-$   & $-$  & $-$ & $-$   & $-$   & $-$ \\
  (50,75] & 57.95 & 8.66 & 20  & 64.71 & 8.19 & 14  & 73.55 & 8.71 & 3   & $-$   & $-$   & $-$ \\
 (75,100] & 80.88 & 8.74 & 5   & 84.31 & 8.61 & 12  & 84.58 & 8.88 & 9   & 93.50 & 10.11 & 9 \\
(100,125] & $-$   & $-$  & $-$ & 108.52& 9.31 & 2   & 109.51& 7.49 & 2   & 115.81& 9.43  & 12 \\
(125,150] & $-$   & $-$  & $-$ & $-$   & $-$  & $-$ & 128.02& 8.36 & 1   & 136.29& 10.33 & 11 \\
(150,175] & $-$   & $-$  & $-$ & $-$   & $-$  & $-$ & $-$   & $-$  & $-$ & 158.09& 11.15 & 4 \\
   $>175$ & $-$   & $-$  & $-$ & $-$   & $-$  & $-$ & $-$   & $-$  & $-$ & 228.48& 9.70  & 9 \\
    Total &  &  & 80  & & & 29   &  &  & 15  &  &  & 45  \\
    (\%)  &  &  & 47.3& & & 17.2 &  &  & 8.9 &  &  & 26.6\\
\hline
\multicolumn{13}{c}{$0.4<[\alpha/{\rm Fe}]\leq 0.7$ dex}\\
\hline
   (0,25] & 20.21 & 3.71 & 5   & $-$   & $-$   & $-$ & $-$  & $-$  & $-$ & $-$    & $-$ & $-$ \\
  (25,50] & 36.24 & 5.76 & 15  & $-$   & $-$   & $-$ & $-$  & $-$  & $-$ & $-$    & $-$ & $-$ \\
  (50,75] & 61.32 & 7.60 & 10  & $-$   & $-$   & $-$ & $-$  & $-$  & $-$ & $-$    & $-$ & $-$ \\
 (50,100] & $-$   & $-$  & $-$ & 77.32 & 10.74 & 3   & $-$  & $-$  & $-$ & $-$    & $-$ & $-$ \\
 (75,100] & $-$   & $-$  & $-$ & $-$   & $-$   & $-$ &79.19 & 5.79 & 3   & $-$    & $-$ & $-$ \\
 (75,150] & $-$   & $-$  & $-$ & $-$   & $-$   & $-$ & $-$  & $-$  & $-$ & 114.95 & 7.05 & 5 \\
    Total &  &  & 30  &  &  & 3   &  &  & 3   &  &  & 5   \\
    (\%)  &  &  & 73.2&  &  & 7.3 &  &  & 7.3 &  &  & 12.2\\
\hline
\end{tabular}
\label{table:3}
}
\end{table*}

Table \ref{table:3} shows that relatively young stars are dominant in the high-probability thin disc population ($0<TD/D \leq 0.1$). However, there are also stars as old as $\tau= 11.15$ Gyr in other populations. The general trend of the age for our sample stars is in agreement with the cited one in the literature, i.e. it increases with increasing of the $[\alpha/{\rm Fe}]$ abundance as well as with total space velocity, and with the four populations in the order of high-probability thin disc, low-probability thin disc, low-probability thick disc, and high-probability thick disc. The distributions of the sample stars in number into the relative probability functions are as follows: $0<TD/D \leq 0.1$: 58646 (84.8$\%$), $0.1<TD/D \leq 1$: 7077 (10.2$\%$), $1<TD/D \leq 10$: 1990 (2.9$\%$), $TD/D>10$: 1435 (2.1$\%$) and nine stars with $0<TD/D \leq 1$ which are led out of discussion.  

\begin{figure}
\centering
\includegraphics[trim=2cm 1cm 1cm 1cm, scale=0.5, angle=0]{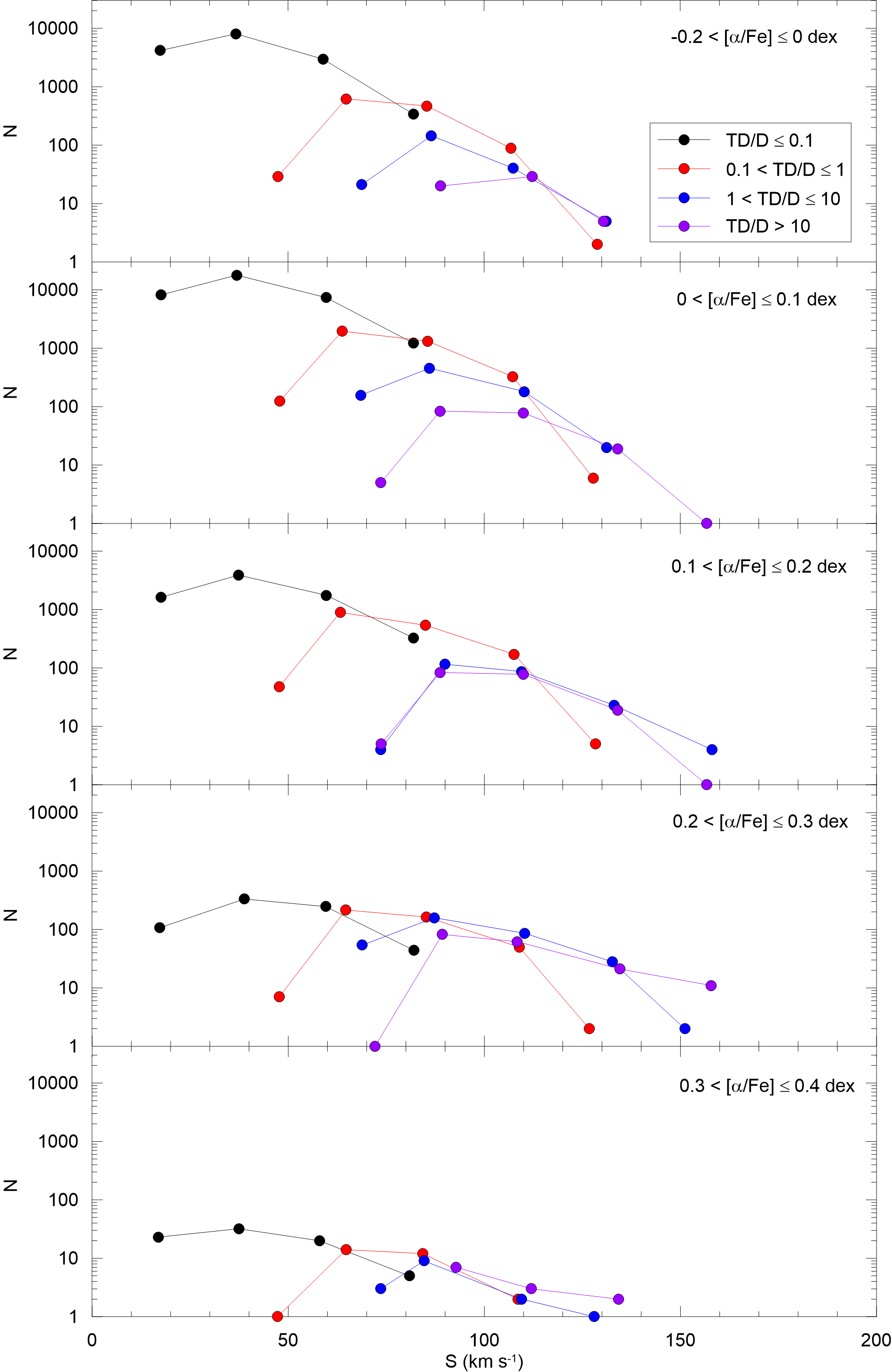}
\caption{The frequency polygons of the total space velocity ($S_{LSR}$) for four relative probability functions, $0<TD/D\leq 0.1$, $0.1<TD/D\leq 1$, $1<TD/D\leq 10$, and $TD/D>10$, for five sub-intervals of $[\alpha /{\rm Fe}]$ as indicated in five panels of this figure.} 
\label{fig5}
\end {figure}

We plotted the frequency polygons of the total space velocities to investigate their trend for each relative probability function of five sub-intervals of $[\alpha/{\rm Fe}]$ where sufficient number of stars are present for discussion, i.e. $-0.2<[\alpha/{\rm Fe}]\leq 0$, $0<[\alpha/{\rm Fe}]\leq 0.1$, $0.1<[\alpha/{\rm Fe}]\leq 0.2$, $0.2<[\alpha/{\rm Fe}]\leq 0.3$, $0.3<[\alpha/{\rm Fe}]\leq 0.4$ dex. Fig. \ref{fig5} confirms the results just cited, i.e. the number of high-probability thin disc stars ($0<TD/D\leq 0.1$) is maximum in all sub-intervals of $[\alpha/{\rm Fe}]$, the lower limit of the total space velocity shifts to larger values as a function of the relative probability $TD/D$, and there is overlapping between the total space velocities of the stars with different relative probabilities, including $0<TD/D\leq 0.1$ and $TD/D>10$ which correspond to the high-probability thin disc and high-probability thick disc, respectively. Then, here again, we can deduce that the total space velocity of a star is not a sufficient parameter to assign its population type.

%Figure 5

We investigated the distributions for the [Fe/H] abundances and ages ($\tau$) of the sample stars for four populations defined by the relative probabilities $TD/D\leq0.1$, $0.1<TD/D \leq1$, $1< TD/D\leq 10$, and $TD/D>10$ without taking into account the $[\alpha/{\rm Fe}]$ abundance of the stars. Table \ref{table:4} shows that the range of the [Fe/H] abundance is rather large, $-1.3< {\rm [Fe/H]}\leq 0.6$ dex. However, the number of [Fe/H]-poor stars are relatively small. Hence, we considered only the stars with $\langle{\rm [Fe/H]}\rangle\pm 2\sigma$, where $\sigma$ is the standard deviation of (all) stars in each population. This constraint limits the range of the [Fe/H] abundance for the populations, in the order just cited, as follows: (-0.6, 0.4], (-0.7, 0.4], (-0.8, 0.4], and (-1.0, 0.2]. The numbers of stars so defined are in boldface in the table. Then, one can say that the lower limit of the range shifts to the poor [Fe/H] abundance values gradually when one goes from the high-probability thin disc to the high-probability thick disc. The same trend holds for the mean of the [Fe/H] abundance estimated for all stars in a population, i.e. -0.08, -0.12, -0.19, and -0.37 dex. The most interesting thing in this table is the existence of an unexpected number of [Fe/H]-poor high-probability thin disc stars, i.e. 1859 stars with ${\rm[Fe/H]}<-0.5$ dex. 

%Table 4
\begin{table}
\setlength{\tabcolsep}{1pt}
\centering
\caption{Frequency distribution for the [Fe/H] abundance of the sample stars in four relative probability intervals. $N$ indicates the number of stars. The total number of stars is 69157. Figures with bold face are explained in the text.}  
\begin{tabular}{cccccc}
\hline
 &   $TD/D \rightarrow$ & [0, 0.1] & (0.1, 1] & (1, 10] & (10, $\infty$) \\
\hline
[Fe/H] int. &	$\langle$[Fe/H]$\rangle$ & 	$N$ &	$N$ & $N$ & $N$ \\
       (dex)& (dex) &  &  &  &  \\
\hline
(-1.3, -1.2]&	-1.25&	1	  &   0	    & 0	    &1\\
(-1.2, -1.1]&	-1.15&	0	  &   1	    & 0	    &2\\
(-1.1, -1.0]&	-1.05&	1	  &   0	    & 2	    &4\\
(-1.0, -0.9]&	-0.95&	4	  &   5	    & 1	    &\bf20\\
(-0.9, -0.8]&	-0.85&	21	  &   14	& 7	    &\bf51\\
(-0.8, -0.7]&	-0.75&	103	  &   54	&\bf26  &\bf67\\
(-0.7, -0.6]&	-0.65&	462	  &\bf162	&\bf88  &\bf163\\
(-0.6, -0.5]&	-0.55& \bf1267&\bf334	&\bf145 &\bf189\\
(-0.5, -0.4]&	-0.45& \bf2976&\bf544	&\bf204 &\bf206\\
(-0.4, -0.3]&	-0.35& \bf5187&\bf703	&\bf236 &\bf189\\
(-0.3, -0.2]&	-0.25& \bf7757&\bf929	&\bf270 &\bf178\\
(-0.2, -0.1]&	-0.15& \bf9528&\bf978	&\bf279 &\bf125\\
(-0.1,  0.0]&	-0.05& \bf9817&\bf905	&\bf253 &\bf102\\
(0.0,  0.1] &    0.05& \bf8900&\bf973	&\bf178 &\bf55\\
(0.1,   0.2]&	 0.15& \bf5959&\bf623	&\bf113 &\bf41\\
(0.2,   0.3]&	 0.25& \bf3587&\bf433	&\bf97  &23\\
(0.3,   0.4]&	 0.35& \bf2272&\bf301	&\bf56  &13\\
(0.4,   0.5]&	 0.45&	747	  &   110	& 33    &6\\
(0.5,   0.6]&	 0.55&	64	  &   10	& 2	    &0\\

\hline
\multicolumn{2}{c}{Total}                  & 58653& 7079  & 1990  & 1435\\
\hline
\multicolumn{2}{c}{$\langle$[Fe/H]$\rangle$ (dex)} &-0.08 &-0.12  & -0.19 &-0.37\\
\hline
\end{tabular}
\label{table:4}
\end{table}

\begin{figure}
\centering
\includegraphics[scale=0.35, angle=0]{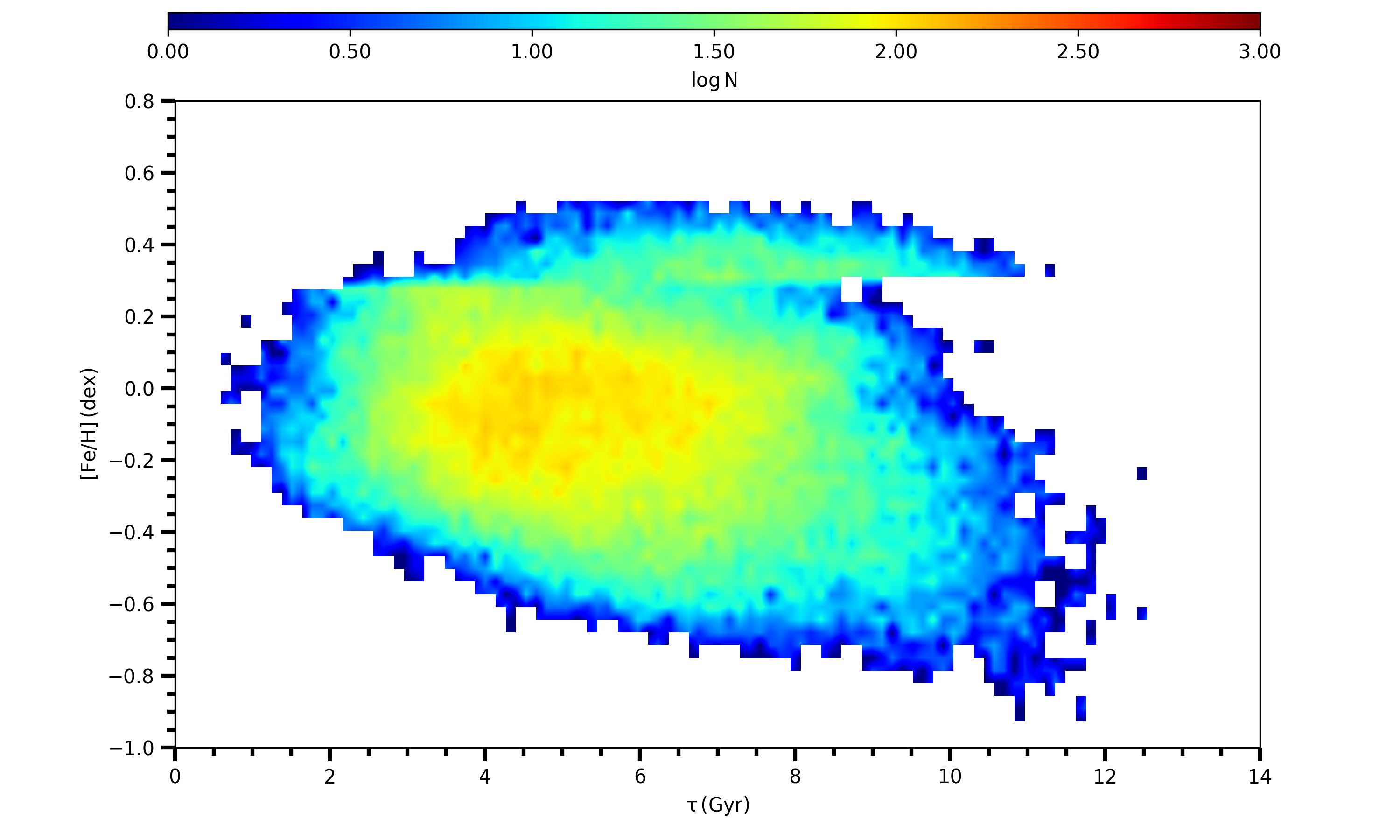}
\caption{Age-metallicity relation for the sample stars.} 
\label{fig6}
\end {figure}

The ages of 75 stars could not be estimated, hence the number of the star sample reduced to 69082. The ranges for the age of the stars in four populations are rather large (Table \ref{table:5}), i.e. (0, 13], (1, 13], (1.5, 13], (1.5, 13] for $TD/D \leq 0.1$, $0.1<TD/D \leq 1$, $1<TD/D \leq 10$, and $TD/D>10$, respectively. However, they reduce as follows when one considers the data for $\langle\tau\rangle \pm2\sigma$: high-probability thin disc: $2<\tau \leq 9$ Gyr, low-probability thin disc: $2.5<\tau \leq 10$ Gyr, low-probability thick disc: $3< \tau \leq 11$ Gyr, and high-probability thick disc: $3.5<\tau \leq 12.5$ Gyr. The numbers of stars so defined are in boldface in the table. The most interesting thing in this table is the existence of 2290 high-probability thin-disc stars older than 9 Gyr where about two dozens of them are coeval with the Universe. Fig. \ref{fig6} shows that the [Fe/H] poor stars, i.e. ${\rm [Fe/H]}<-0.5$ dex, are dominant by the old stars. Hence, one can say that most of the stars with unexpected [Fe/H] abundances (the 1859 stars mentioned in the foregoing paragraph) and 2290 stars just mentioned, are identical (we shall discuss this problem in \S 4).

%Table 5
\begin{table}
\setlength{\tabcolsep}{1pt}
\centering
\caption{Frequency distribution for the age ($\tau$) of the sample stars in four relative probability intervals. $N$ indicates the number of stars. Figures with bold face are explained in the text.}  
\begin{tabular}{cccccc}
\hline
 &   $TD/D \rightarrow$ & [0, 0.1] & (0.1, 1] & (1, 10] & (10, $\infty$) \\
\hline
$\tau$ int. &	$\langle \tau \rangle$ & $N$ &	$N$ & $N$ & $N$ \\
(Gyr)& (Gyr) &  &  &  &  \\
\hline
(0, 0.5]	&0.25	&21	 &$-$	    & $-$    & $-$\\
(0.5, 1.0]	&0.75	&62	 &$-$	    & $-$    & $-$\\
(1.0, 1.5]	&1.25	&219     & 4	    & $-$    & $-$\\
(1.5, 2.0]	&1.75	&594     & 19	    & 4	     & 4\\
(2.0, 2.5]	&2.25	&\bf1282 & 32	    & 12     &	5\\
(2.5, 3.0]	&2.75	&\bf2414 & \bf90    & 15     &	10\\
(3.0, 3.5]	&3.25	&\bf3851 & \bf190   & \bf41  &	15\\
(3.5, 4.0]	&3.75	&\bf5139 & \bf340   & \bf62  &	\bf24\\
(4.0, 4.5]	&4.25	&\bf5690 & \bf464   & \bf87  &	\bf38\\
(4.5, 5.0]	&4.75	&\bf5928 & \bf548   & \bf105 &	\bf44\\
(5.0, 5.5]	&5.25	&\bf5751 & \bf698   & \bf129 &	\bf65\\
(5.5, 6.0]	&5.75	&\bf5418 & \bf689   & \bf157 &	\bf66\\
(6.0, 6.5]	&6.25	&\bf5113 & \bf705   & \bf181 &	\bf82\\
(6.5, 7.0]	&6.75	&\bf4601 & \bf691   & \bf200 &	\bf92\\
(7.0, 7.5]	&7.25	&\bf3764 & \bf588   & \bf187 &	\bf93\\
(7.5, 8.0]	&7.75	&\bf2901 & \bf521   & \bf168 &	\bf101\\
(8.0, 8.5]	&8.25	&\bf2101 & \bf415   & \bf145 &	\bf101\\
(8.5, 9.0]	&8.75	&\bf1456 & \bf326   & \bf115 &	\bf103\\
(9.0, 9.5]	&9.25	& 933    & \bf282   & \bf115 &	\bf143\\
(9.5, 10.0]	&9.75	& 596    & \bf194   & \bf100 &	\bf127\\
(10.0, 10.5]	&10.25	& 374    & 118	    & \bf65  &	\bf96\\
(10.5, 11.0]	&10.75	& 212    & 76	    & \bf54  &	\bf90\\
(11.0, 11.5]	&11.25	& 95	 & 31	    & 20     &	\bf56\\
(11.5, 12.0]	&11.75	& 49	 & 23	    & 16     &	\bf50\\
(12.0, 12.5]	&12.25	& 17	 & 15	    & 7	     & \bf15\\
(12.5, 13.0]	&12.75	& 14	 & 13	    & 4	     & 6\\
\hline
\multicolumn{2}{c}{Total}               &   58595& 7072& 1989& 1426\\
\hline
\multicolumn{2}{c}{$\langle \tau \rangle$ (Gyr)} &5.49 &6.45 &7.07 &8.13\\
\hline
\end{tabular}
\label{table:5}
\end{table}

\begin{figure}
\centering
\includegraphics[scale=0.7, angle=0]{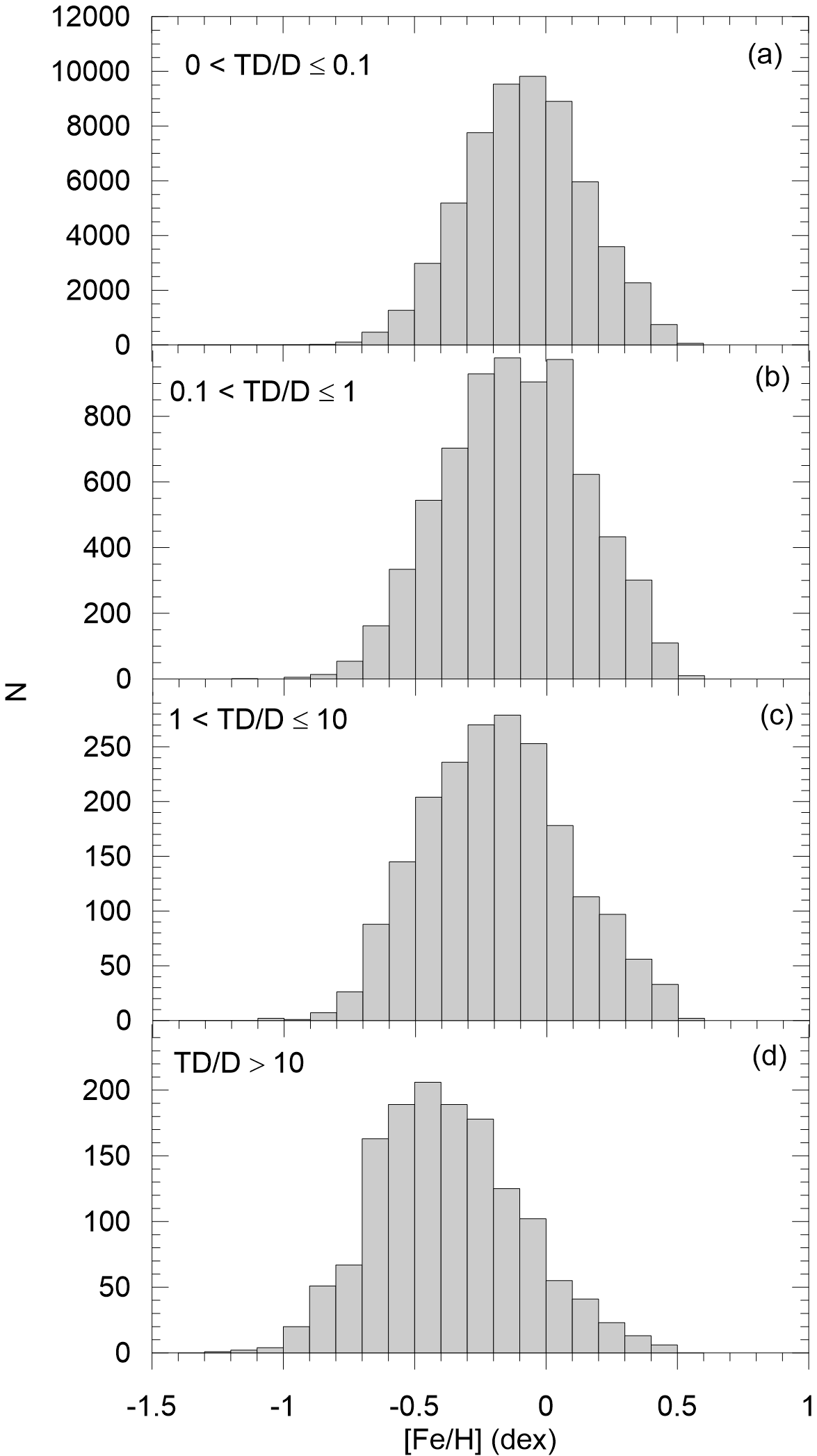}
\caption{Histograms for the [Fe/H] abundance of four relative probability functions as indicated in the panels (a)-(d).} 
\label{fig7}
\end {figure}

\begin{figure}
\centering
\includegraphics[scale=0.7, angle=0]{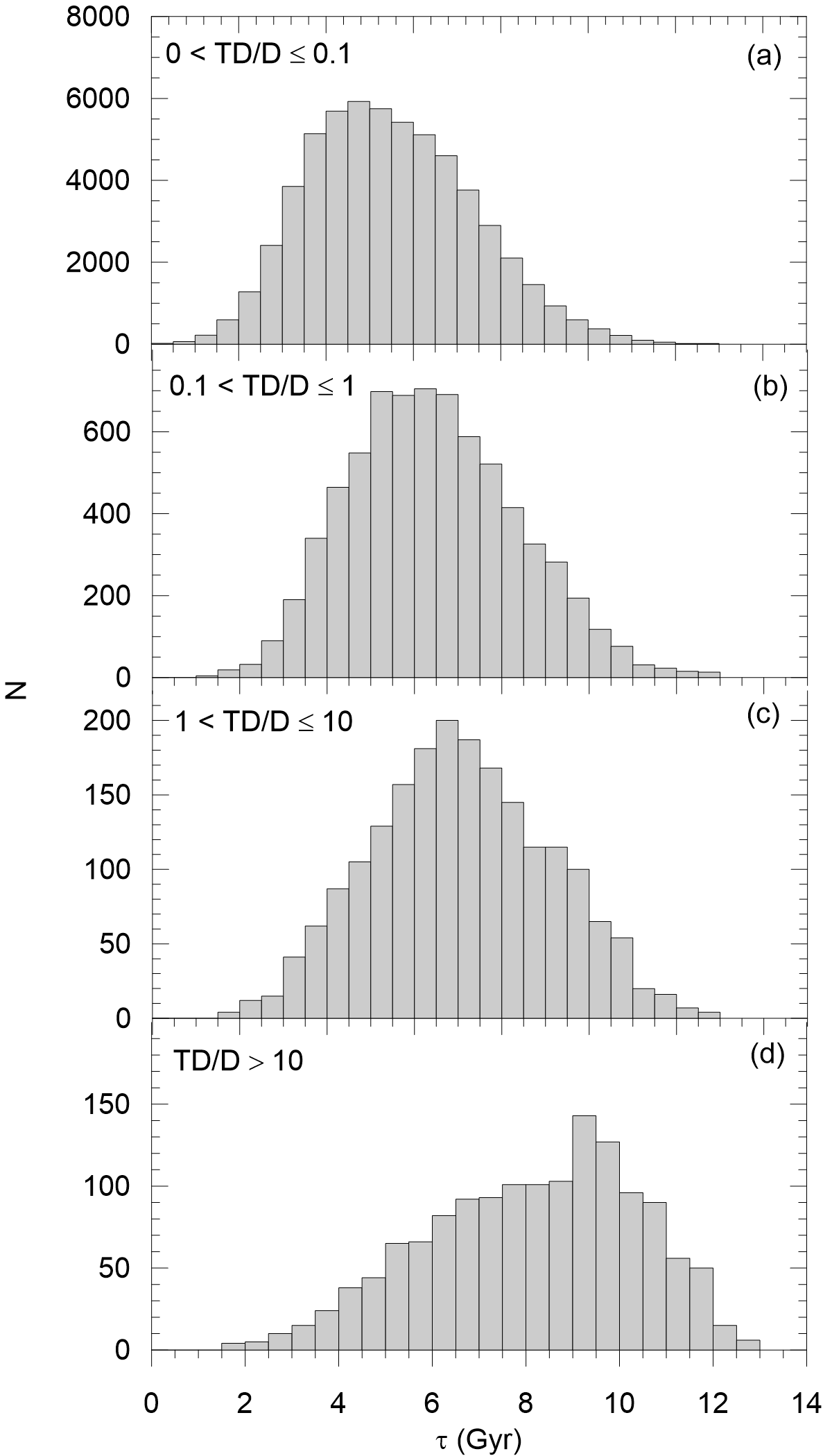}
\caption{Histograms for the age ($\tau$) of four relative probability functions as indicated in the panels (a)-(d).} 
\label{fig8}
\end {figure}

The distributions of the [Fe/H] abundances and ages for four populations are plotted in Fig. \ref{fig7} and Fig. \ref{fig8}, respectively. The mode of the [Fe/H] distribution shifts to the metal-poor direction, gradually, when one goes from the high-probability thin disc to low-probability thin disc, low-probability thick disc and high-probability thick disc. While the mode of the age distribution shifts to the old ages, gradually, when one goes from the one  probability function to the other one in the order just mentioned. 

%Figure 6

%Figure 7

%Figure 8

%Figure 9
\begin{figure}
\centering
\includegraphics[scale=0.42, angle=0]{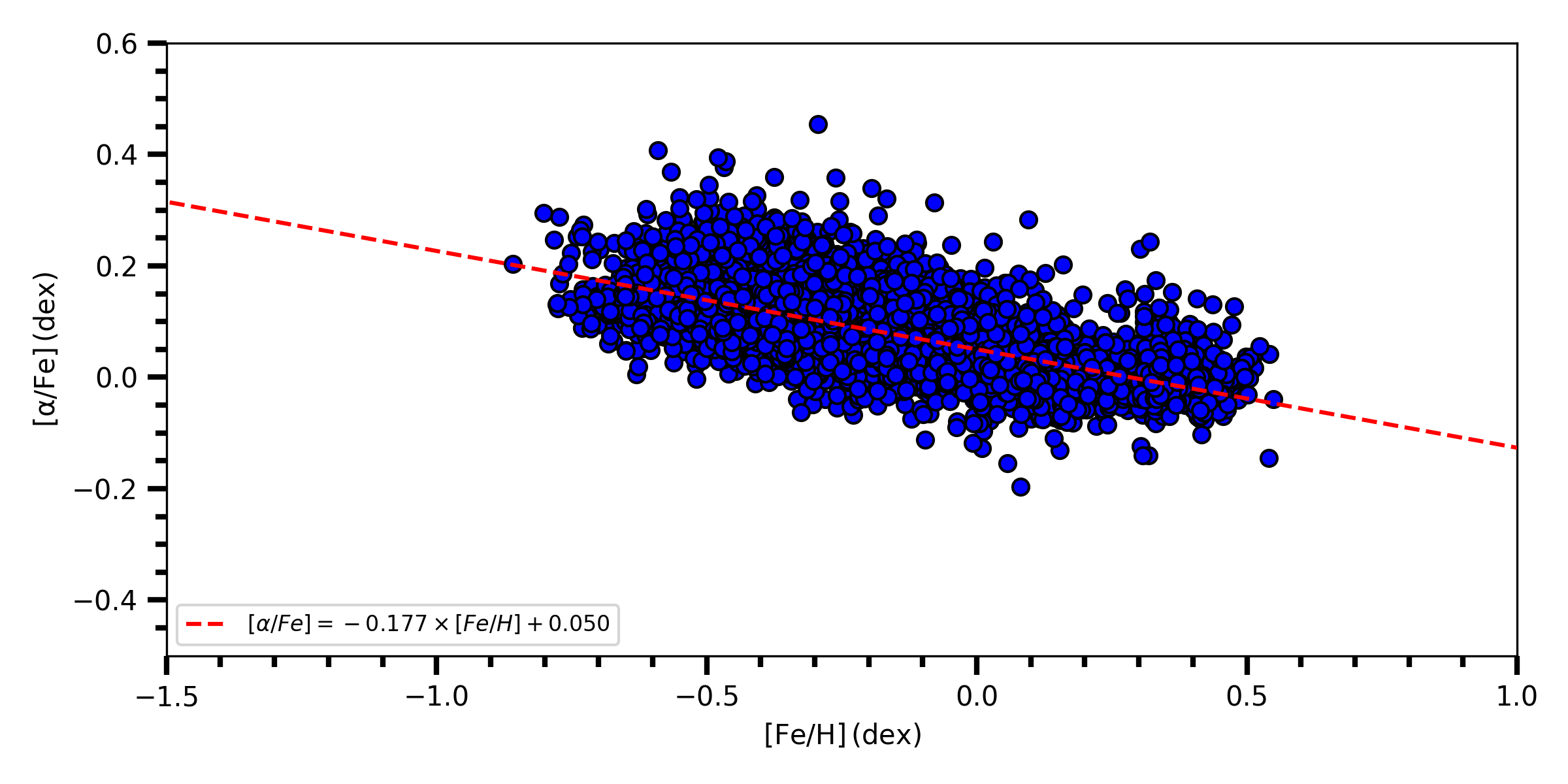}
\caption{[$\alpha$/Fe]$\times$[Fe/H] diagram for the sample stars with age $8<\tau\leq 9$ Gyr used for estimation of the abundance [$\alpha$/Fe]=0.14 dex which separates the stars into $\alpha$-rich and $\alpha$-poor categories.}
\label{fig9}
\end {figure}

\subsection{Investigation of the sample stars via spectroscopic procedure: Distribution of the [$\alpha$/Fe] abundance in terms of metallicity and age}
\begin{figure}
\centering
\includegraphics[scale=0.42, angle=0]{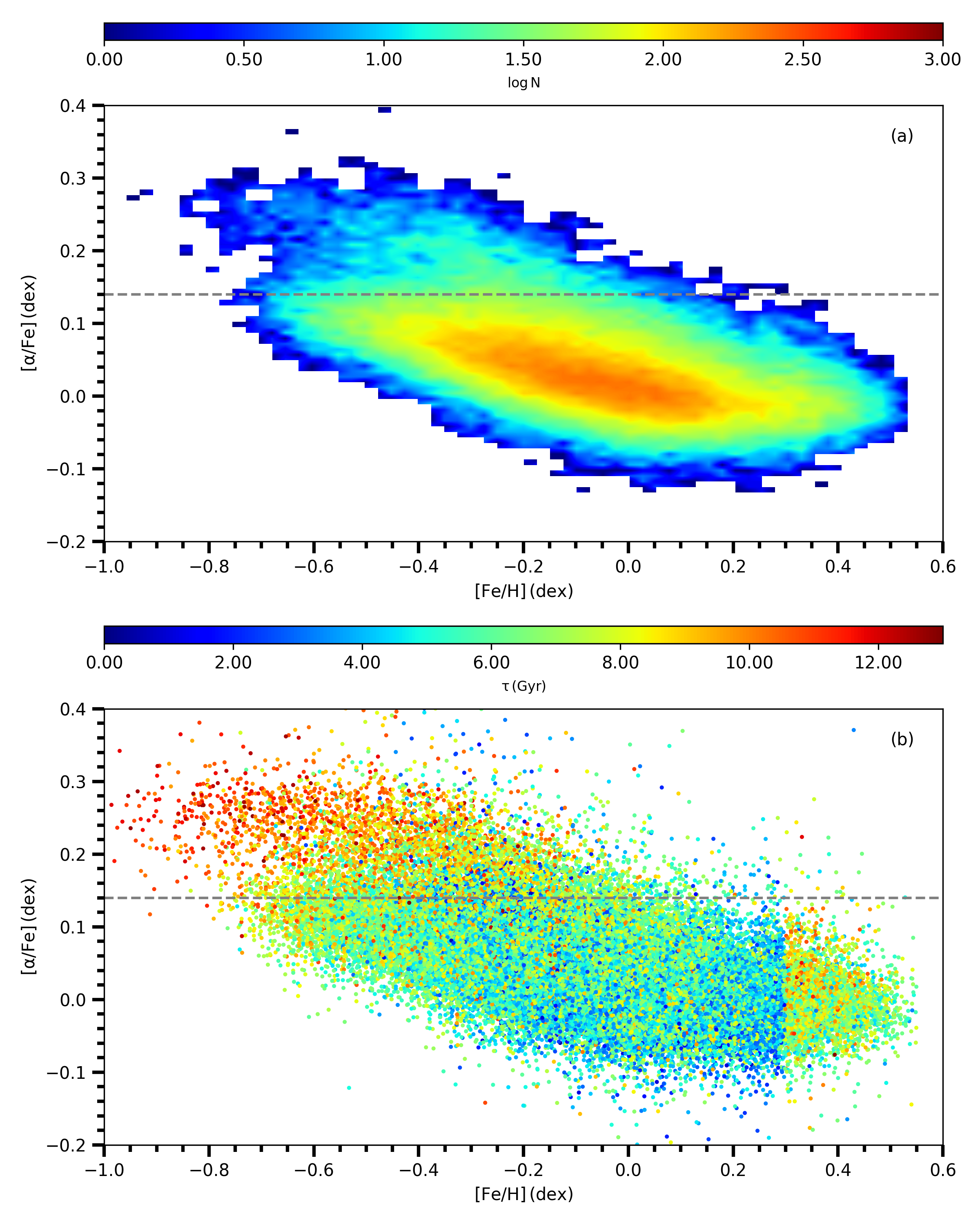}
\caption{[$\alpha$/Fe]$\times$[Fe/H] diagram for the sample stars and the horizontal line [$\alpha$/Fe]=0.14 dex which separates them into $\alpha$-rich and $\alpha$-poor categories  (a), and the same diagram with colour coded for age (b).}
\label{fig10}
\end {figure}

In this section, we separated the sample stars into two categories via their [$\alpha$/Fe] and [Fe/H] abundances, and age. Following \citet{Bland-Hawthorn19}, we used the ``$\alpha$-rich'' and ``$\alpha$-poor'' terminology instead of thin disc and thick disc. Also, we used the results obtained by \citet{Haywood19}. Different procedures have been used in the literature to distinguish the $\alpha$-rich and $\alpha$-poor (or high-$\alpha$ and low-$\alpha$) stars in the [$\alpha$/Fe]$\times$[Fe/H] plane \citep{Adibekyan12, Bensby14, Hayden15, Bland-Hawthorn19}. Here, we avoided of using kinematical parameters, instead we combined the [$\alpha$/Fe] abundance with the [Fe/H] one and age to distinguish the two star categories just mentioned. Following \citet{Haywood19}, we adopted the age $\tau=8$ Gyr as the boundary for separating the stars into $\alpha$-rich and $\alpha$-poor categories. Actually, the [$\alpha$/Fe]$\times \tau$ diagram in (their) Fig. 1 represents a combination of two different distributions, one with $\alpha$-rich and old stars ($\tau>8$ Gyr) and another one with $\alpha$-poor and young ones ($\tau<8$ Gyr). This procedure is also in agreement with the trend of our data in the panel (c) of Fig. 3, i.e. the stars with age $\tau>8$ Gyr are richer in [$\alpha$/Fe] than the younger ones. We plotted the sample stars with $8<\tau\leq 9$ Gyr in the [$\alpha$/Fe]$\times$[Fe/H] diagram (Fig. \ref{fig9}) and fitted them to a linear equation. Then, we estimated the [$\alpha$/Fe] abundances corresponding to the [Fe/H] abundances -0.6, -0.5 an -0.4 dex which cover the $\alpha$-rich stars in the [$\alpha$/Fe]$\times$[Fe/H] diagram (Fig. \ref{fig10}). The corresponding [$\alpha$/Fe] abundances and their mean are as follows: [$\alpha$/Fe]=0.16, 0.14, 0.12, and $\langle[\alpha/{\rm Fe}]\rangle=0.14$ dex. Thus, we adopted the abundance [$\alpha$/Fe]=0.14 dex as the boundary separating the sample stars in our study as $\alpha$-rich and $\alpha$-poor stars. This procedure is similar to the one in \citet{Buder19} who showed that stars of the high-$\alpha$ sequence are older ($>8$ Gyr) than stars in the low-$\alpha$ sequence with the iron abundances $-0.7<[{\rm Fe/H}]<+0.5$ dex, and applied their finding to the metal-rich stars which become indistinguishable in the [$\alpha$/Fe]$\times$[Fe/H] diagram, as mentioned in the \S 1. Thus, the number of $\alpha$-rich and $\alpha$-poor dwarfs observed in the GALAH survey are $N_1$=5778 and $N_2$=63379, respectively. A bimodal distribution in the [$\alpha$/Fe]$\times$[Fe/H] diagram fails due to the less number of $\alpha$-rich stars relative to $\alpha$-poor ones, i.e. 9\%.               

%Figure 10

\section{Summary and Discussion}
We used the spectroscopic and astrometric data provided from the GALAH DR2 \citep{Buder18} and  {\it Gaia} DR2 \citep{Gaia18}, respectively, for a large sample of stars to investigate the behaviour of the $[\alpha/{\rm Fe}]$ abundances via two procedures, i.e. kinematically and spectroscopically. With the kinematical procedure, we investigated the distribution of the $[\alpha/{\rm Fe}]$ abundances into the thin disc and thick disc populations in terms of total space velocity, $[{\rm Fe/H}]$ abundance, and age. We applied the following constrains to the total sample of stars (342682 stars) to obtain a star sample of F-G type dwarfs with accurate trigonometric parallaxes: $5250 \leq T_{eff} \leq 7160$ K, $\log g \geq 4$ (cm s$^{-2}$), $\sigma_{\varpi}/\varpi \leq0.05$, and $G\leq14$ mag. The number of our sample stars is 69157. The radial velocities are provided from the GALAH DR2, while the proper motions and trigonometric parallaxes are taken from the {\it Gaia} DR2. Radial velocities, proper motions and distances are combined to obtain space velocity components for the sample stars which are calculated with \citet{Johnson87} standard algorithms. Also, the differential rotation correction is applied to the $U$ and $V$ space velocity components. 

We separated the sample stars into seven sub-intervals of $[\alpha/{\rm Fe}]$, i.e. $[\alpha/{\rm Fe}]\leq -0.2$, $-0.2<[\alpha /{\rm Fe}]\leq 0$, $0<[\alpha/{\rm Fe}]\leq 0.1$, $0.1<[\alpha/{\rm Fe}]\leq 0.2$, $0.2<[\alpha/{\rm Fe}]\leq 0.3$, $0.3<[\alpha/{\rm Fe}]\leq 0.4$, and $0.4<[\alpha/{\rm Fe}]\leq 0.7$ dex, and investigated the distributions of the stars therein for four populations, i.e. high-probability thin disc ($0<TD/D\leq 0.1$), low-probability thin disc ($0.1<TD/D\leq1$), low-probability thick disc ($1<TD/D\leq 10$), and high-probability thick disc ($TD/D>10$). We omitted the discussion of nine stars in the sub-interval $[\alpha/{\rm Fe}]\leq -0.2$ dex. In other sub-intervals of the $[\alpha/{\rm Fe}]$, the total space velocity interval for the high-probability thin disc stars is $0<S_{LSR}\leq 100$ km s$^{-1}$, while its lower and upper limits shift gradually to higher values when one goes to low-probability thin disc, low-probability thick disc and high-probability thick disc stars. Also, the number of high-probability thin disc stars are maximum in all sub-intervals of $[\alpha/{\rm Fe}]$ abundance. This finding is in contradiction with our expectation, because one expects the number of high-probability thick disc stars (not the thin disc ones) to dominate in relatively high $[\alpha/{\rm Fe}]$ sub-intervals, such as $0.3<[\alpha/{\rm Fe}]\leq 0.4$, and $0.4<[\alpha/{\rm Fe}]\leq 0.7$ dex. However, the percentage of the high-probability thick disc stars are relatively higher in two high [$\alpha/{\rm Fe}]$ sub-intervals, i.e. 28.4$\%$ and 26.6$\%$ in the sub-intervals $0.2<[\alpha/{\rm Fe}]\leq 0.3$ and $0.3<[\alpha/{\rm Fe}]\leq 0.4$ dex, respectively. Another unexpected result is the domination of the high-probability thin disc stars in the high $0.4<[\alpha/{\rm Fe}]\leq 0.7$ dex interval, though the number of stars is only 41. Also, the low total space velocities of these stars, $20.21 \leq S_{LSR} \leq 61.32$ km s$^{-1}$, and their young ages, $3.71 \leq \tau \leq 7.60$ Gyr, confirm the argument that they belong to the high-probability thin disc population. However, the domination of the high-probability thin disc stars in all sub-intervals of $[\alpha/{\rm Fe}]$ including the high one, $0.4<[\alpha/{\rm Fe}]\leq 0.7$ dex, can not be generalized. This result is due to the two constrains applied in this study, i.e. the limitation of the apparent magnitude ($V<14$ mag) and the intermediate Galactic latitudes of the sample stars. Another study which would cover a sample of high-Galactic latitude, and apparently faint stars would be rather useful. In this case, we expect a distribution for the sample stars where the high-probability thick disc stars would dominate in the high $[\alpha/{\rm Fe}]$ abundances.

\begin{figure}
\centering
\includegraphics[scale=0.8, angle=0]{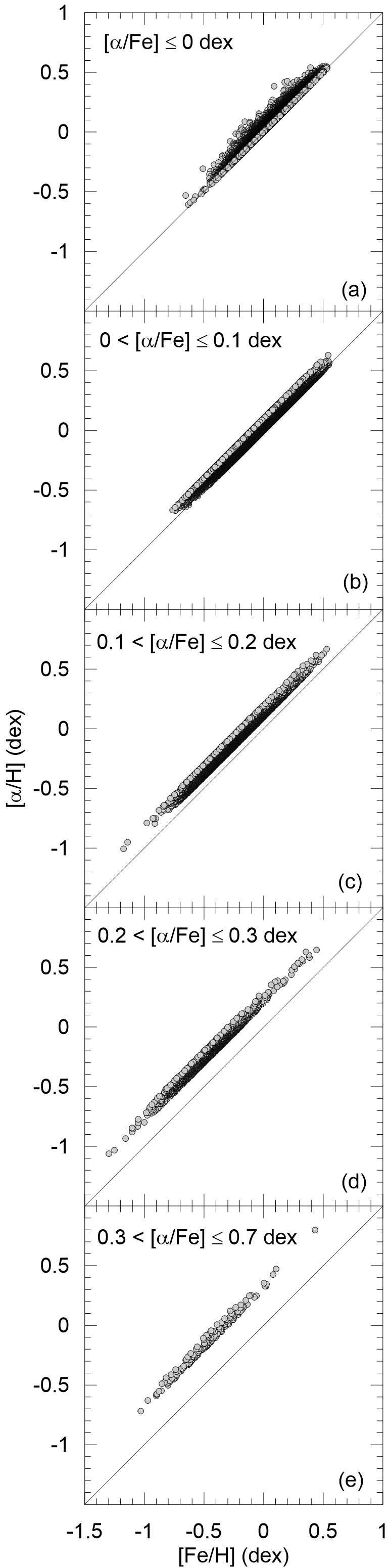}
\caption{Distributions of the $[\alpha /{\rm H}]$ abundance against [Fe/H] abundance for five $[\alpha/{\rm Fe}]$ intervals as indicated in the panels (a)-(e). Notice the deviation of the distribution from one-to-one straight line in the direction of increasing of the $[\alpha /{\rm Fe}]$ abundance.} 
\label{fig11}
\end {figure}

The distributions of the sample stars into the sub-intervals $-0.2<[\alpha/{\rm Fe}]\leq 0$, $0<[\alpha/{\rm Fe}]\leq 0.1$, $0.1<[\alpha/{\rm Fe}]\leq 0.2$, $0.2<[\alpha/{\rm Fe}]\leq 0.3$, $0.3<[\alpha/{\rm Fe}]\leq 0.4$, and $0.4<[\alpha/{\rm Fe}]\leq 0.7$ dex show that each of these intervals covers the stars of different populations with different percentages, however. That is, many stars of different populations share the same $[\alpha/{\rm Fe}]$ abundance. For instance, 731 (35$\%$) high-probability thin disc stars and 594 (28.4$\%$) high-probability thick disc stars share the abundance interval $0.2<[\alpha/{\rm Fe}]\leq0.3$ dex, in Table \ref{table:3}. Additionally, 453 (4.4$\%$) high-probability thick disc stars share the lower interval $0.1<[\alpha/{\rm Fe}]\leq0.2$ dex with 7584 (73.5$\%$) high-probability thin stars, in the same table. The interpretation of the low $[\alpha/{\rm Fe}]$ abundance is generally based on the rich iron-abundance of the (thin disc) stars. According to this argument, the iron-abundance becomes rich in the interstellar medium via the Type Ia supernovae, much later than the formation of the $\alpha$-elements in the same medium. Then, one should expect rich $[\alpha/{\rm Fe}]$ abundance for old/[Fe/H]-poor stars, while the young/[Fe/H]-rich ones would be [$\alpha/$Fe]-poor. However, our results cited in this paragraph show that relatively young and iron rich (high-probability thin disc) stars and relatively older and poorer in iron abundance (high-probability thick disc) stars share the same $[\alpha/{\rm Fe}]$ abundance. This result can be explained as follows: the richness of the $[\alpha/{\rm Fe}]$ abundance depends on the richness of the $[\alpha/{\rm H}]$ and [Fe/H] abundances. If these abundances are compatible, then the star is $[\alpha/{\rm Fe}]$-poor while (relatively) a large difference between [$\alpha/$H] and [Fe/H] indicates an [$\alpha/$Fe]-rich star (we kindly remind to the readers that $[\alpha/{\rm Fe}]=[\alpha/{\rm H}]-[{\rm Fe/H}]$). We separated our sample stars into five $[\alpha/{\rm Fe}]$ intervals, i.e. $[\alpha/{\rm Fe}]\leq 0$, $0<[\alpha/{\rm Fe}]\leq 0.1$, $0.1<[\alpha/{\rm Fe}]\leq 0.2$, $0.2<[\alpha/{\rm Fe}]\leq 0.3$, and $0.3<[\alpha/{\rm Fe}]\leq 0.7$ dex and plotted their $[\alpha/{\rm H}]$ abundances versus [Fe/H] ones in Fig. \ref{fig11} for clarification of our argument. The distributions of the $\alpha$-poor stars ($[\alpha/{\rm Fe}]\leq 0$ and $0<[\alpha/{\rm Fe}]\leq 0.1$ dex) are almost superposed with the one-to-one straight line, while it deviates gradually from this line when one goes to richer $\alpha$-abundances. The degree of the deviation corresponds to the difference between the $[\alpha/{\rm H}]$ and [Fe/H] abundances. Each panel covers [Fe/H]-rich stars as well as stars with $-0.7<[{\rm Fe/H}]\leq 0$ dex. Also, the [Fe/H] abundances of the stars with richer $\alpha$-abundances ($[\alpha/{\rm Fe}]>0.1$ dex) extends beyond ${\rm [Fe/H]}=-1$ dex. In summary, we can say that stars with different [Fe/H] abundances may share the same $[\alpha/{\rm Fe}]$ abundance, and to be [Fe/H] rich is a sufficient -but not a necessary- condition for a star to be $[\alpha/{\rm Fe}]$ poor. 

%Figure 11

The domains of the [Fe/H] abundances and the ages for the sample stars are rather large, i.e. $-1.3<{\rm [Fe/H]}\leq 0.6$ dex and $0<\tau \leq 13$ Gyr, respectively. However, when we consider only the $\langle {\rm[Fe/H]}\rangle \pm 2\sigma$ and $\langle\tau \rangle \pm 2\sigma$ data, the cited domains reduce as follows for four populations: high-probability thin disc: $-0.6<{\rm [Fe/H]} \leq 0.4$ dex, $2<\tau\leq 9$ Gyr; low-probability thin disc: $-0.7<{\rm [Fe/H]} \leq 0.4$ dex, $2.5<\tau \leq 10$ Gyr; low-probability thick disc: $-0.8<{\rm [Fe/H]}\leq 0.4$ dex, $3<\tau \leq 11$ Gyr; and high-probability thick disc: $-1<{\rm [Fe/H]}\leq 0.2$ dex, $3.5<\tau \leq 12.5$ Gyr where $\sigma$ is the standard deviation of the data for the cited population, and where $\langle {\rm [Fe/H]}\rangle \pm 2\sigma$ and $\langle \tau\rangle \pm 2\sigma$ cover 95.45$\%$ of the data in question. This procedure eliminates the young/metal-poor stars which are classified as high-probability thick disc/high-probability thin disc stars. The most interesting thing in the distribution of the iron abundances and ages of the sample stars is the existence of a large number of metal poor -and old- stars in the high-probability thin disc population, i.e. 1859 stars with ${\rm [Fe/H]} \leq -0.5$ dex, and 2250 stars with $\tau> 9$ Gyr. According to the age-metallicity relation of the sample stars (Fig. \ref{fig6}), there is an important number of stars which share the [Fe/H] abundances and ages just cited. That is, we have a sub-sample of high-probability thin disc stars with compatible [Fe/H] abundance and age with high-probability thick disc stars. This finding confirms the argument that the metal-poor thin disc stars were formed at the same epoch of the thick disc stars with $8<\tau<10$ Gyr in a different location of the Galaxy \citep[cf.][]{Haywood13}. Also, the upper limit [Fe/H]=0.2 dex claimed for the high-probability thick disc is in agreement with the argument of \citet{Haywood19} who stated that the thick disc enriched up to the solar value nine Gyr ago. We used the mean of the [$\alpha$/Fe] abundance, i.e. $\langle[\alpha/{\rm Fe}]\rangle=0.14$ dex, for the sample stars with age $8<\tau\leq9$ Gyr and $[{\rm Fe/H}]\leq-0.4$ dex to separate them into $\alpha$-rich and $\alpha$-poor categories. This value is compatible with the one in \citet{Bland-Hawthorn19} claimed for the giants observed in the GALAH survey (their Fig. 3).

\begin{figure*}[ht]
\centering
\includegraphics[scale=0.5, angle=0]{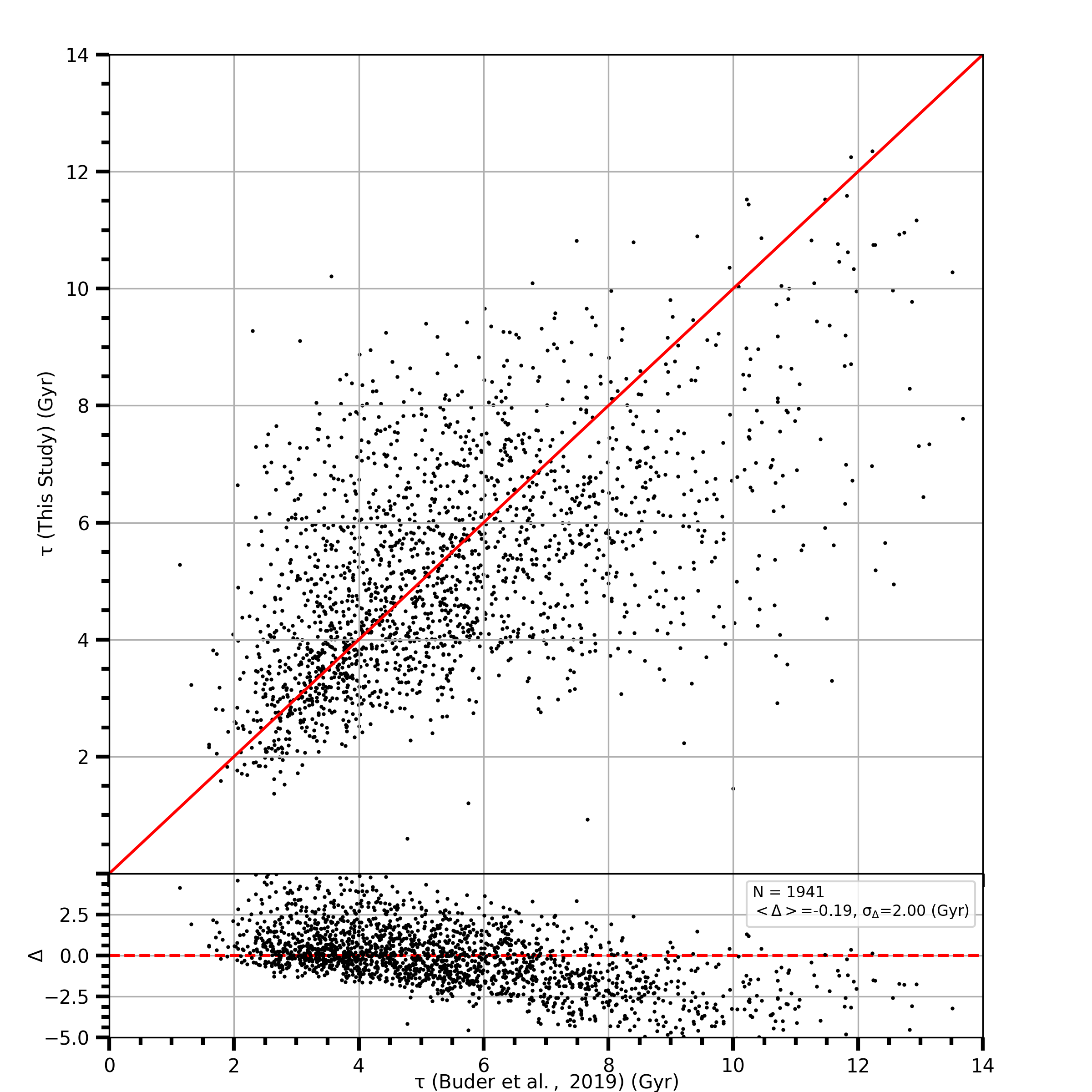}
\caption{Ages estimated in our study (ordinate axes) versus the ones in \citet{Buder19}, (upper panel) and distribution of the age differences in two studies in terms of ages in \citet{Buder19}, (lower panel).} 
\label{fig12}
\end {figure*}

\begin{figure*}[ht]
\centering
\includegraphics[width=10cm,height=10cm, keepaspectratio]{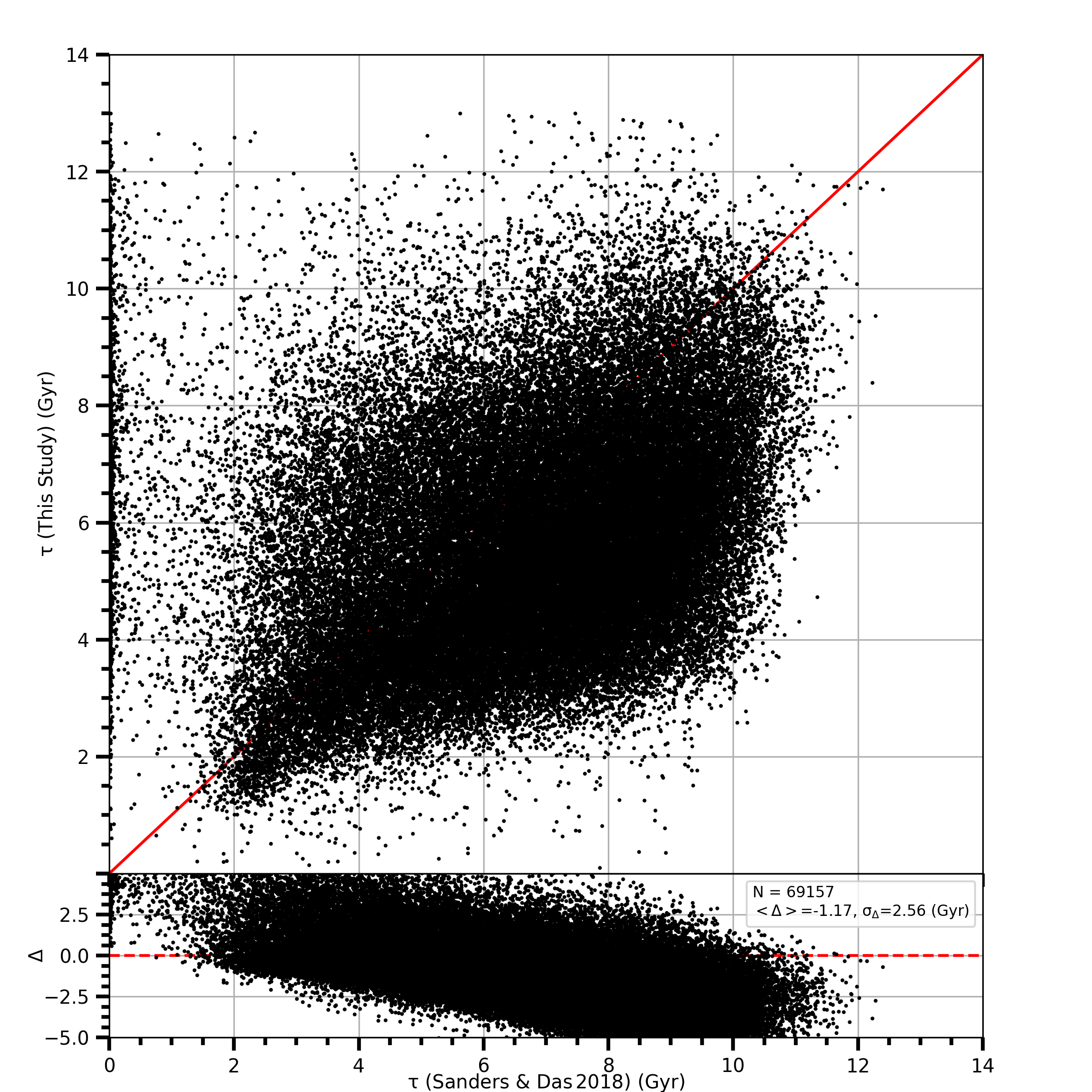}
\caption{Comparison of the ages estimated in our study and in \citet{Sanders18} via the procedure used in Fig. \ref{fig12}.} 
\label{fig13}
\end {figure*}

\begin{figure*}
\centering
\includegraphics[width=10cm,height=10cm, keepaspectratio]{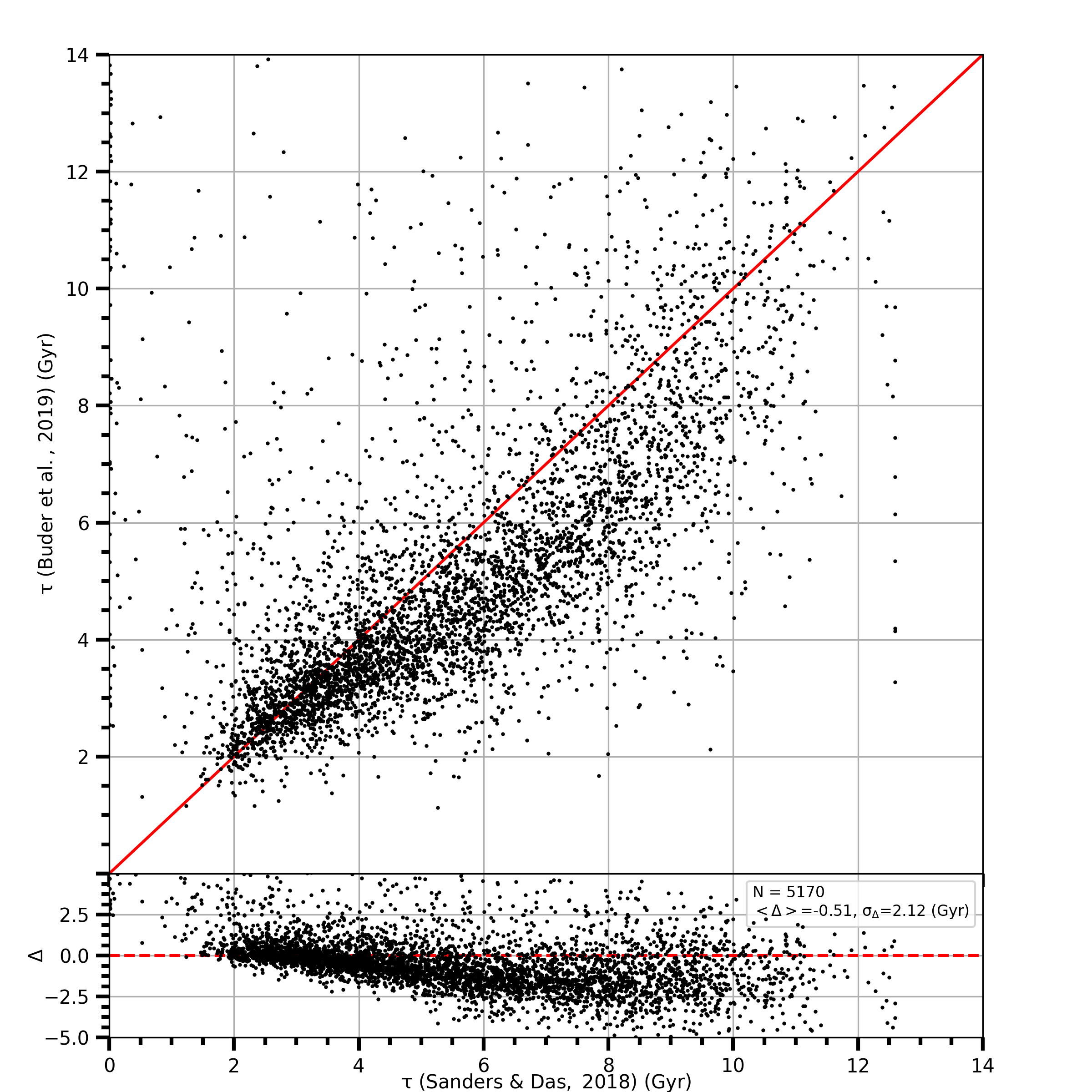}
\caption{Comparison of the ages in \citet{Buder19} and \citet{Sanders18} via the procedure used in Fig. \ref{fig12}.} 
\label{fig14}
\end {figure*}

We compared the ages estimated in our study with the ones in \citet{Buder19} and \citet{Sanders18}. The common stars in our study and in \citet{Buder19} is only 1941. The mean of the differences between two sets of ages and the corresponding standard deviation are 0.19 and 2 Gyr, respectively. One can see an agreement in Fig. \ref{fig12} for the two sets of stars with age $\tau<6$ Gyr, while there is a deviation for older ages which probably is due to the different isochrones used in different studies, i.e. PARSEC in our study and Darmouth in \citet{Buder19}. The mean of the differences between the ages in our study and in \citet{Sanders18} for 69157 stars and the corresponding standard deviation are a bit larger than the former ones, i.e. 1.17 and 2.56 Gyr. As can be seen in Fig. \ref{fig13}, the difference between the ages of two studies is larger for the older stars. We compared also the ages in \citet{Buder19} and \citet{Sanders18} just to check their agreement (Fig. \ref{fig14}). The mean of the differences between the ages of 5170 common stars in two studies and the corresponding standard deviation are 0.50 and 2.14 Gyr, i.e. a bit larger than the ones 0.19 and 2 Gyr cited for the comparison of the ages in our study with the ones in \citet{Buder19}. Hence, one can say that the ages estimated in our study are as accurate as the ones appeared in the literature. 

\begin{figure*}
\centering
\includegraphics[width=16cm,height=16cm, keepaspectratio]{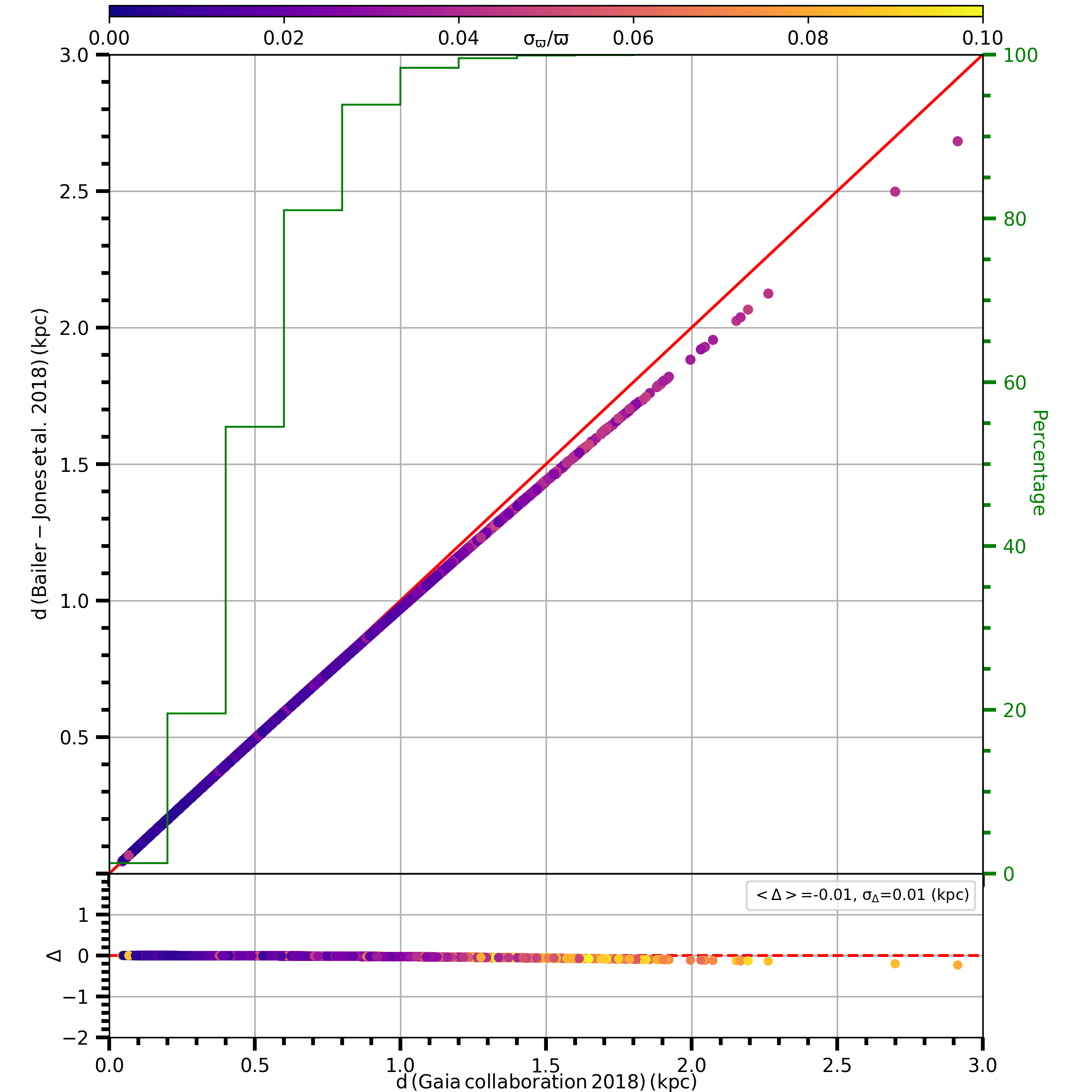}
\caption{Comparison of the distances for the sample stars estimated via two procedures, i.e. {\it Gaia} DR2 trigonometric parallaxes (abscissa axes) and geometrically (ordinate axes). 90\% of the stars are within one kpc distance and they fit with the one-to-one straight line.} 
\label{fig15}
\end {figure*}

We tested the accuracy of the distances estimated in our study, by using the {\it Gaia} DR2 trigonometric parallaxes, by comparison them with the ones estimated via the geometric procedure which is usually used in the literature \citep{Bailer-Jones18}. The mean of the differences between the distances estimated by two procedures and the corresponding standard deviation are both 0.01 kpc. Also, one can see in Fig. \ref{fig15} that most of the data fit with the one-to-one straight line, while only a few percent of them deviate from this line. Actually, the cumulative frequency curve shows that 90\% of the data overlap with the curve just cited.

The final comparison is carried out between our data and the ones in a recent study, \citet{Casali19}, in the ([$\alpha/{\rm Fe}$], [Fe/H], age) space. Our Fig. 10b and their Fig. 13 have almost the same trend. However the distribution in \citet{Casali19} shows a rather more clear distinction between the $\alpha$-rich/thick disc and $\alpha$-poor/thin disc populations. Hence, we deduce that the GALAH data we used is perhaps too noisy to show more than a general outline of the  underlying relationships.     

%Figure 12

%Figure 13

%Figure 14

%Figure 15

\section{Acknowledgments}
We thank to Gerry Gilmore for his valuable comments which improved our paper. This research has made use of NASA's (National Aeronautics and Space Administration) Astrophysics Data System and the SIMBAD Astronomical Database, operated at CDS, Strasbourg, France and NASA/IPAC Infrared Science Archive, which is operated by the Jet Propulsion Laboratory, California Institute of Technology, under contract with the National Aeronautics and Space Administration. This work has made use of data from  the European Space Agency (ESA) mission {\it Gaia} (\mbox{https://www.cosmos.esa.int/gaia}), processed by the {\it Gaia} Data Processing and Analysis Consortium (DPAC, \mbox{https://www.cosmos.esa.int/web/gaia/dpac/consortium}). Funding for the DPAC has been provided by national institutions, in particular the institutions participating in the {\it Gaia} Multilateral Agreement.

\end{document}